# Characterization and Evaluation of Carbonaceous Materials from the Hydrothermal Carbonization of Waste Pharmaceuticals

*Carla M. Ndoun[1] and Samuel A. Darko[2*]*

[1]CHA Consulting 1016 Spring Villas Point, Winter Springs, FL 32708

[2]School of Arts and Sciences, Florida Memorial University, 15800 NW 42[nd]Ave, Miami Gardens, FL 33054

*Corresponding author: Phone: (305) 626-3157; Fax: email: samuel.darko@fmuniv.edu

## ABSTRACT

In this paper, we report herein, the conversion of waste prescription and non-prescription pharmaceuticals into carbonaceous materials. The hydrothermal carbonization (HTC) of the pharmaceuticals was carried out at temperatures of 180, 230 and 275 $^0$C in closed reactors for 6, 12 and 24 hours, respectively. The main products from the carbonization process were in the solids, liquids and gas phases. The resulting hydrochars were shown to be very functionalized with a high degree of aromaticity and high carbon content (between 55% to 65%). The adsorptive capacity of the hydrochars to remove $Pb^{2+}$ ions from an aqueous system was evaluated and compared to that of analytical reagent activated carbon (AR-AC) through batch adsorptive tests. The effect of contact time on batch adsorption experiments with an initial $Pb^{2+}$ concentration of 50 mg/L was also evaluated. The results indicated that PH24_230 has a better adsorption capacity when compared to AR-AC; achieving over 97% removal of $Pb^{2+}$ after 60 minutes. The batch adsorption studies were best described by the pseudo-second order kinetic model with coefficient of regression ($R^2$) values above 0.99. Also, slow pyrolysis experiments were carried out to evaluate the difference in solid yields, char heating value and surface area. Pyrochar yields were slightly higher, while heating values were one order of magnitude lower when compared to hydrochars. The pyrolysis conducted at 700°C led to the pyrochar with the highest value of the surface area (63.15 $m^2/g$). The study shows that valuable products can be generated successfully from the hydrothermal carbonization of waste pharmaceuticals.

**Graphical Abstract**

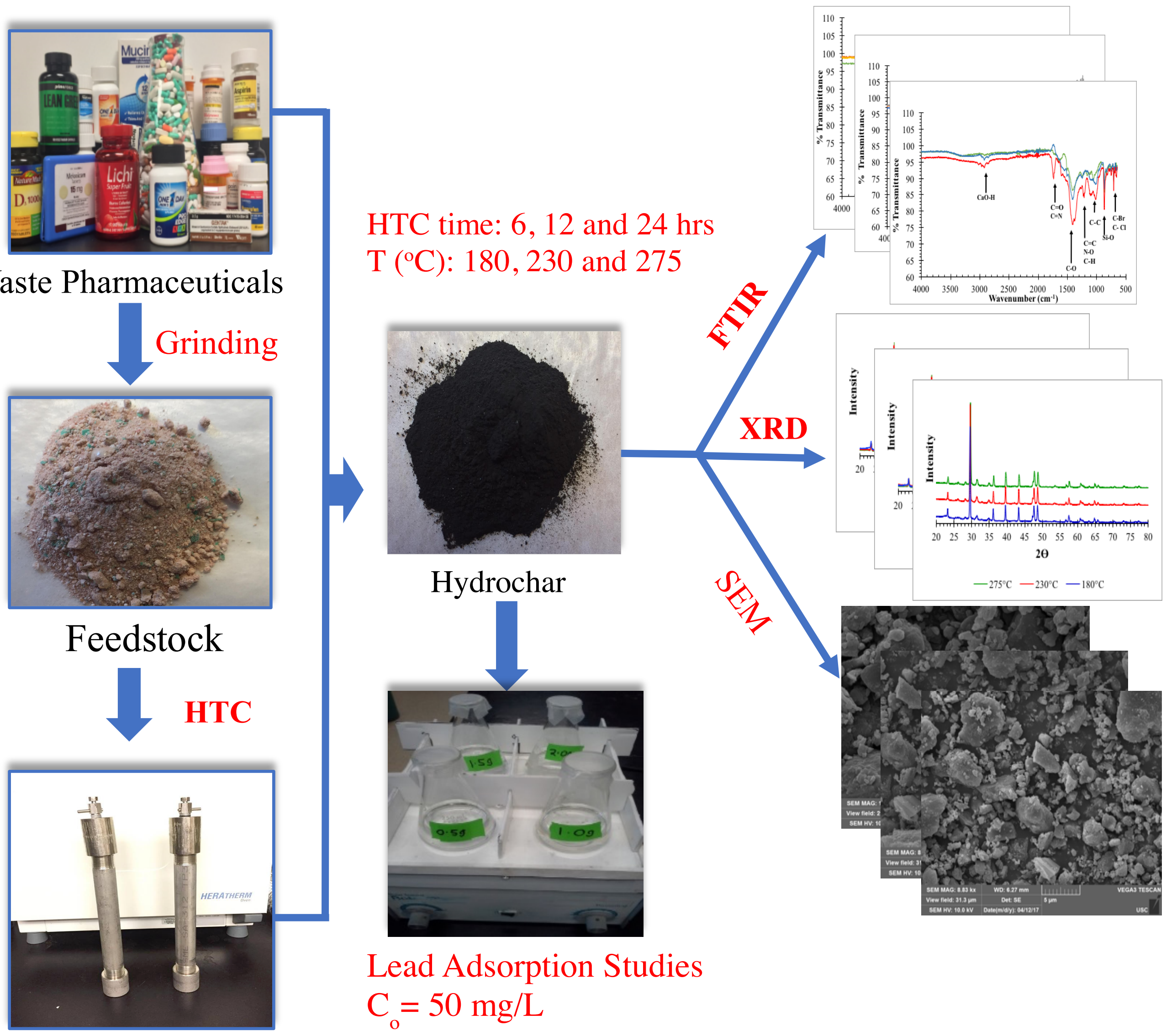

## 1.0 Introduction

Hydrothermal Carbonization (HTC) has gained attention as a viable means for the conversion of biomass and wastes streams into a carbon-rich, energy dense, value-added solid materials referred to as hydrochar (Titirici et al., 2007; Berge et al., 2011; Hwang et al., 2012; Li et al., 2013). HTC is described as a thermochemical process for converting an organic feedstock into value-added products, at moderate temperatures (180-350 $^0$C) and pressures (2-10 MPa) in the presence of liquid water (Funke and Felix, 2010; Libra et al., 2011).

HTC has proven to be a promising technology for treatment and conversion of diverse biomass feedstock such as algal biomass (Broch et al., 2013), food waste (Berge et al., 2015; Chowdhury et al., 2014, Sabio et al., 2015, Fernandez et al., 2015), municipal waste streams (Berge et al., 2011), woody and herbaceous feedstock (Mazzoleni et al., 2013), plant material (Titirici et al., 2007, Wang et al., 2012), textile waste (Lin et. al., 2016), seafood waste (Kannan et al., 2015) to obtain energy-dense bio-crude (Ehumen et al., 2011; Akhtar and Amin, 2011; Heilmann et al., 2010, 2011a,b; Hoekman et al., 2011; Titirici et al., 2007; Xiao et al., 2012), soil ammendment to increease soil fertility and crop yields (Du et al., 2012; Kleinert et al., 2009; Libra et al., 2011; Rillig et al., 2010), carbon materials that could be activated to work as an adsorbent for the decontamination of water systems (Ledesma et al., 2015, Rajasekhar et al., 2015), photoluminescent polymer nanodots (Lui et al., 2012), for catalytic (Xiao et al. 2012), electronic and biological applications and much more. HTC is carried out at relatively low temperatures (180–350 C) in the presence of water under autogenous pressures (Titirici et al., 2007; Berge et al., 2011, Hoekeman et al., 2011; Libra et al., 2011), and does not require the drying of traditionally wet feedstocks for waste conversion. Also, a large percentage of the carbon in the original feedstock remains integrated within the hydrochar, ultimately minimizing greenhouse gas emission (Parshetti et al., 2013). HTC of waste materials may require fewer solids processing/ treatment (such as mechanical dewatering of biosolids) and handling (hydrochar is sterilized).

To date, out of all the possible waste streams that have been studied as feedstock for hydrothermal carbonization, an area that has not been fully elucidated is a study evaluating expired or unwanted pharmaceuticals. Pharmaceutical products refer to a group of chemicals used for diagnosis, treatment, or prevention of diseases and health conditions. These products are classified into two categories: Over the counter (OTC) which can be bought at supermarkets and pharmacies without any restriction and Prescription Only Medicines (POM) which, to be obtained, require physician's prescription (Gualtero et al., 2005). The United States leads the world in pharmaceuticals products, accounting for 39% of the world's production. As a result, hospitals and long-term care centers flush approximately 250 million pounds of unused pharmaceuticals down the drain each year (Arleen, 2012). This has motivated a search for ways to use expired or unwanted pharmaceuticals waste.

Currently, the most common methods used to discard waste pharmaceuticals (tablets, syrups, capsules, drops, inhalers, suppositories) which are flushing, landfilling, chemical decomposition and incineration have proven to have a degree of effectiveness. However, the above methods have several economic and environmental drawbacks. These include; high cost for the operation and maintenance of these methods, emission of methane which is which has 20–25 times stronger warming effect than $CO_2$ on a molecular basis, thereby contributing significantly to climate change (Nahman et al., 2012) and drug diversion which involves the transfer of any legally prescribed controlled substance from the individual for whom it was prescribed to another person for any illicit use. Furthermore, water treatment plants cannot fully eliminate the chemicals found in pharmaceuticals leading to trace amounts found in water bodies.  Alternatively, Plasma Thermal Destruction and Recovery (PTDR) for the destruction of pharmaceuticals has been explored. This technology uses heat generated by plasma torches in an anaerobic environment, to dissociate the molecules that make up the organic portion of the pharmaceuticals (Capote and Ripes, 2009). Although this method has proven to be an alternative to landfilling and incineration, it remains expensive due to the high temperatures (over 800 $^{0}C$) required to breakdown organics into gas and chemicals involved (hydrogen peroxide, potassium chlorate) which are usually added to reform the dissociated elements of the waste into a synthesis gas, consisting mainly of Carbon Monoxide (CO) and Hydrogen ($H_2$).

The need for more cost effective and environmentally acceptable method for the disposal of waste pharmaceutical is currently under research. HTC as a conversion technique has the potential to overcome many of the challenges associated with the conventional disposal methods and chemical treatment of expired or unwanted pharmaceuticals. The HTC process is comparatively low emission and a hazardous waste free process because it uses water as the sole reaction medium under pressure and heat (Hoekman et al., 2011). Studies have also identified potential environmental benefits associated with using HTC in this manner, such as a reduction in greenhouse gas emissions and lower energy requirements for the conversion of wet feedstocks when compared to more traditional waste conversion processes (Titirici et al., 2007; Ramke et al., 2009; Berge et al., 2011; Falco et al., 2011; Román et al., 2013). Furthermore, hydrochars obtain from the HTC of biomass differ from biochars in that they are generally less aromatic, consisting of mostly alkyl moieties (Cao et al., 2009). Work conducted by Sun et al. (2011, 2012) showed that hydrothermally produced hydrochars have the potential to adsorb a greater number of polar and nonpolar organics in water more effectively than biochar, demonstrating the use of hydrochars as a useful adsorbent for decontamination of wastewaters.

In the search for environmentally suitable methods to manage and treat expired or unwanted pharmaceuticals and also to evaluate the carbon content of such materials for environmental and energy-related applications, the overarching objectives of the present work were thus (i) to investigate the potential of HTC treatment to upgrade pharmaceuticals, by increasing its carbon content and thus its heating value (ii) study the effects of two variables; temperature and reaction time on the yield and characteristics of the hydrochars obtained, (iii) examine the microstructure and chemical composition of the proudced hydrochar (iv) evaluate the potential application of the hydrochars as an adsorbent for the removal of lead. Lead was selected due to it's extensive use in paint and batteries and it's reported detrimental effects in humans such as irritability, developmental delays, abdominal pain and neurologic changes,.

## 2. Materials and Methods

### 2.1 Raw Materials

The feedstock used in this study constituted of 14 Over the counter (OTC) and 10 Prescription only medicine (POM) (Fig. S1 and Table S1) that were donated by a local health care facility as their recommended shelf-life had passed or they were no longer required by the user. Expired pharmaceuticals are drugs which potency and safety decline after the expiration date has passed. The pharmaceuticals were ground using a coffee grinder to reduce their surface area and increase homogeneity to ease the carbonization process. The moisture content of the pharmaceuticals was determined by weighing a known amount of the feedstock and drying it in the oven at 80°C until a consistent weight was obtained during the consecutive weighing. The elemental composition, microstructure, surface area and pore size, higher heating value (HHV), and functional groups present in the feedstock were determined using the Energy-Dispersive X-ray Spectroscopy (EDX), Scanning Electron Microscopy (SEM), Brunauer–Emmett–Teller analysis (BET), Bomb Calorimetry, Fourier Transform Infrared Spectroscopy (FTIR) respectively.

**2.2 Batch Hydrochar Production**

HTC of the pharmaceuticals was conducted in 160-mL stainless steel tubular reactors (2.54 cm i.d., 25.4 cm long, MSC, Inc.) fitted with gas-sampling valves (Swagelock, Inc.). Precisely 20g of feedstock was added to 20 ml of deionized water to maintain a 1:1 biomass-to-water ratio in the reactor. HTC at three different temperatures; 180, 230 and 275 $^0$C with reaction times of 6, 12 or 24 hours was carried out in this study. Observed reaction pressures were essentially those of water alone at the respective temperatures, and gaseous products were not analyzed. After the reaction time was reached, the reactors were removed from the oven and subsequently placed in a cold-water bath until the reactors reached room temperature. The cooling stage which took about 10-15 minutes, was not counted as reaction time. After cooling, the gaseous product was discharged in a fume hood, and the solid phase was separated from the liquid by vacuum filtration and subsequently dried at 80 $^0$C to remove residual moisture. Dried hydrochars were placed into plastic bags and stored for further physico-chemical analyses. Each individual experiment was conducted in triplicates. Hydrochars were denoted using the following nomenclature; PH(t)_T where PH stands for pharmaceuticals, t for reaction time and T for the carbonization temperature. Following procedures previously described (Reza et al., 2016; Yan et al., 2010), the mass yields of the HTC reactions was performed considering deionized water and feedstock as input and dried hydrochar and the process liquid and gas as outputs. The unaccountable amount that is not in the solid or the process liquid was assumed to be gas.

### 2.3 Batch Pyrochar Production

Slow pyrolysis batch experiments for pyrochar production were carried out in a LECO TGA 701 thermo-gravimetric analyzer. 1.5 g of crushed feedstock was placed in a ceramic crucible and heated to the desired reaction temperatures of 180, 230, 275, 500 and 700 °C, with a heating rate of 10°C/min under a constant $N_2$ flow (5 l/min). Once the reaction temperature was reached, the samples were maintained at this temperature for 1 hour. In the following cooling step, the nitrogen flow was adjusted to 3.5 l/min until 150°C. Similarly, to hydrochars, pyrochars were denoted using the following nomenclature: PH-TGA(t)_T.

### 2.4 Characterization of Solid Hydrochar

#### 2.4.1 Fourier Transform Infrared Spectroscopy (FTIR)

A PerkinElmer (Spectrum 100) FTIR was used to determine the functional groups present in the feedstock and the obtained hydrochar samples. It is interesting to note that hydrochars with higher degrees of carbonization are highly desirable, particularly in applications such as fuel development. However, in this study hydrochars with increased functionality was important since the produced material was to be applied for adsorption studies (Parshetti et al. 2014). The FTIR spectra of the feedstock and dried and ground hydrochars were obtained using a Perkin Elmer spectrum FTIR spectrometer. About 0.5-1mg of the dry powdered samples were scanned from 4000 cm-1 to 400 cm-1 with a resolution of 4 cm-11.

#### 2.4.2 X-ray powder diffraction (XRD)

X-ray diffraction (XRD) analysis was used to detect crystallinity of a sample, the crystal structures and their lattice parameters, crystallite size and strain, all information that can be vital in material characterization and quality control. XRD was performed on both the raw and hydrochar samples using a computer controlled X-ray diffractometer. All powdered samples were characterized using a benchtop X-ray diffractometer, Rigaku Miniflex II, equipped with a silicon slit detector and a CuKα beam. Diffraction patterns were collected from 2ϴ = 20° to 80° with 0.02° step size and 3°/min scan rate. XRD patterns with crystalline phase (significant peaks) were analyzed with Rigaku's PDXL 2.0 software for searching candidate phases in the ICSD/ICDD database.

**2.4.3 Scanning Electron Microscopy (SEM)**

The surface features of both the raw feedstock and the hydrochar were examined by SEM imaging with Hitachi TM-3000 (Tokyo, Japan) scanning electron microscope. Varying resolution from 50x to 2000x was used to analyze the samples. Imaging was done in high vacuum mode at an accelerating voltage of 20kV, using secondary electrons. The signals that are derived from electron-sample interactions reveal information about the sample including external morphology (texture), chemical composition, and crystalline structure and orientation of materials making up the sample.

**2.4.4 Energy-Dispersive X-ray Spectroscopy (EDX)**

Elemental analysis was performed to qualitatively detect the major elements within the raw feedstock and hydrochar sample. The carbon content for each sample was used to perform carbon densification studies. EDX analysis was performed by using a Zeiss DSM 962 (Carl Zeiss, Oberkochen, Germany).

**2.4.5 Higher heating value**

The Higher heating value of the feedstock and selected hydrochars and pyrochars were measured with the calorimeter LECO AC500, according to UNI 14918. Each sample was weighted and put in a crucible; the crucible was then inserted in the bomb calorimeter and a nickel wire connected with two electrodes. The bomb was sealed and pressurized with pure oxygen (grade 5.0) up to 30 bar and then inserted in 2 liters of water inside the calorimeter. After 3 minutes for equilibrating temperature, combustion was triggered, and the higher heating value was obtained after 5 minutes. The thermocouple of the analyzer has a sensibility of 0.0001°C. To ensure complete combustion; the samples were burnt with benzoic acid, per the certified standard, and was also used for the calibration of the thermocouple. All measurements were conducted in duplicates.

### 2.4.6 Brunauer–Emmett–Teller analysis (BET)

Brunauer–Emmett–Teller (BET) analysis for surface area, total pore volume, and average pore diameter was performed on the raw feedstock, on selected hydrochars, and the pyrochars. The analysis was conducted with the Quantachrome NOVA 2200E BET analyzer, according to ISO 9277. Before the analysis, the samples were heated in an oven for 48 h (100°C for the hydrochars, 160°C for the pyrochars) to remove water and some volatiles, and then they were degassed under vacuum at 160°C for 4 h in the analyzer. After the degassing, the samples were weighted and analyzed with pure nitrogen (grade 5.0) at 77 K. Adsorption, and desorption isotherm curves were recorded for pressure ratios between 0.005 and 1 ($P_0$ = 100146.43 Pa). Surface area was measured in the adsorption isotherm curve for pressure ratios between 0.005 and 0.30. The total pore volume was calculated in the adsorption isotherm curve for a pressure ratio of 1. Instrument calibration was verified with certified standards (alumina pellets and activated carbon spheres). The average pore diameter was determined according to eq. 1:

$$Average\ pore\ diameter\ (nm) = 4(V_p/A_s) \quad (1)$$

where $V_p$ is the total pore volume and $A_s$ is the surface area.

## 2.6 Batch Adsorption Experiments.

Lead (II) nitrate was purchased from Fisher Scientific, New Jersey was utilized as the model contaminant without further purification. The selection of this lead was based on its prevalence in water, particularly, the severe case of lead posing experienced in Flint, Michigan. Lead was maintained at an initial concentration ($C_o$) of 50 mg/L. The concentration of Pb was analyzed using the Agilent Technologies atomic absorption spectrophotometer (200 series AA). A sample of contaminated water containing $Pb^{2+}$ ions is atomized, usually by a flame or graphite furnace, and dispersed into the light. A detector measures the amount of absorption in the sample and compares it to a reference with a known concentration of the lead in question to determine its concentration in the sample.

Batch adsorption experiments were conducted to investigate the efficiencies of the hydrochars (PH24_180, PH24_230, PH24_275, PH12_180, PH12_230, PH12_275, PH6_180, PH6_230, and PH6_275) to determine the most optimal adsorbent for the removal of $Pb^{2+}$ from simulated wastewater. The effect of contact time was studied at an initial concentration of 50mg/L for $Pb^{2+}$ and the adsorptive capabilities of the hydrochars were determined and compared to the analytical reagent activated carbon (AR-AC) as reference. The activated carbon with a surface area of 490 $m^2/g$ was purchased from Calgon Carbon (USA) was used without any further preparation. A 100-mL aliquot of the contaminant solution was added to a 250 mL Erlenmeyer flask containing 1.0g of adsorbent (hydrochar and AR-AC) at room temperature and agitated by a shaker at 160 rpm. The total contact time was 60 minutes, and 15 mL of aliquots were collected after every 10 minutes and filtered using a vacuum flask and filter paper to determine the change in concentration of the $Pb^{2+}$ ions. Each experiment was run in triplicates, and the mean values were reported accordingly. The adsorption capacity at time *t* was calculated using the equation

$$q_t = \frac{(C_o - C_t)V}{M} \quad (2)$$

where $q_t$ (mg/g) is the adsorption capacity at any time $t$, $C_0$ and $C_t$ are the initial concentration (mg/L) of lead ions in solution and the concentration at any time $t$, $V$ (L) is the volume, and $M$ (g) is the weight of the adsorbent.

## 3. Results and Discussions

### 3.1 Hydrochar Yield

The yield from the HTC of pharmaceuticals at various HTC conditions is presented in Table 1. This table demonstrates the average mass in, of the feedstock into the reactors. The mass of the hydrochar recovered, the mass of liquid and mass of the gas from the triplicates experiments conducted at three different temperatures (180, 230 and 275 $^0$C) and reaction times (6, 12 and 24 hours) are reported as well. The yield analysis indicated that the hydrochar production decreased with an increase in reaction temperature for either 6, 12 or 24 hours' reaction times with hydrochar production as low as 13.55g for PH24_275. From table 1, an increase in the process temperature from 180 $^0$C to 230 $^0$C to 275 $^0$C resulted in a decrease in the mass of the hydrochar from 19.47 g to 18.18 g to 14.20 g respectively.

Table 2 and fig. 1 describe the percent yield of the hydrochar, liquid and gas phases for the three different HTC temperatures and reaction times. The hydrochar production decreases with an increase in reaction temperature. The hydrochar yield obtained from the hydrothermal carbonization are in the 41 to 48 wt% range at 180 $^0$C, 35-45 wt% at 230 $^0$C and 33-35 wt% at 275 $^0$C. A similar trend was observed during the HTC treatment of several different biomass available in literature (Castello et al., 2013; Lui et al., 2012; Libra et al., 2011; Kang et al., 2012; Hoekman et al., 2011). This outcome was due to a larger overall decomposition of biomass material, or because of the presence of carbon in the pharmaceuticals matter, which progressively reacted and diffused out. It can also be seen from fig. 1 and table 2 that the hydrochar yields decrease as the time is increased from 6 to 24 hours. At 180 $^0$C, the hydrochar yield decreases from 48.48% to 43.15% to 41.18% when the time increases from 6 to 12 to 24 hours respectively. This similar trend is observed when keeping the temperature constant (at 230 and 275 $^0$C) and increasing the reaction time from 6 hours to 12 hours and finally to 24 hours. At all three temperatures, it can be seen that after 12 hours of treatment, the hydrochar yield does not change significantly. This indicates that any major transformations and structural rearrangements do occur in the first 12 hours, after which the products became structurally stable. The same trend was observed during the production of hydrochars from sewage sludge by He et al. (2013).

The liquid product reported in tables 1 and 2 includes four different liquids: liquid present in the pharmaceutical feedstock (moisture content =8.18%), liquids collected by vacuum filtration after the HTC experiments, residual liquids evaporated during the drying of the hydrochar in the oven and liquid soaked by the filter paper. The first step of the HTC reaction is hydrolysis that requires water to react with the feedstock to facilitate the breaking of bonds enabling the formation of new ones (Reza et al., 2014). The effect of time and temperature in the production of HTC process liquid is also visible as the net liquid out increases with reaction time at any specific HTC temperature and 48-61% of the liquid is present after the HTC experiment.

### 3.2 Slow pyrolysis of the feedstock: mass yields and de-volatilization behavior

The slow pyrolysis of the feedstock performed at 500 and 700°C with 1-hour holding time led to a solid yield of 60 and 55 weight %, respectively. In fig. 2, the derivative of the TGA versus furnace temperature is reported. Mass devolatilization firstly occurred regarding moisture and light organics removal up to 150°C, where a small peak is observed. Further devolatilization took place at 200°C, but the majority of mass loss occurred at 330°C. Another consistent mass reduction happened at nearly 470°C, where heavier organic compounds were volatilized. After 500°C no other de-volatilization peaks were observed, and mass loss remained constant until up to 700°C.

### 3.3 Physico-Chemical Characteristics of the Hydrochar

#### 3.3.1 FTIR Spectra of Hydrochar

To understand the chemical changes in the hydrochar during the HTC at different reaction times and temperatures, FTIR spectroscopy was performed for the feedstock (fig. 3a) along with all hydrochar samples. The infrared spectra of the feedstock and hydrochar samples revealed their complex chemical bond structure consisting of mixture of mineral and organic matter.

From fig. 3a, the peak observed at 2260 $cm^{-1}$ in the feedstock which indicates the presence of the C≡N stretch (Nitrile). This group disappears as the pharmaceuticals were carbonized at different temperatures and reaction times, respectively (fig. 3b, 3c, and 3d). The loss of functional groups is due to the diverse thermal decomposition upon hydrothermal synthesis (Sabio et al., 2015). A clear shift in the spectrum was observed when moving from the feedstock (fig. 3a) to the hydrochars (fig. 3b, 3c, and 3d) as the intensity of the functional groups, bands and peaks increased.

The FTIR spectra of the feedstock and hydrochars obtained at 6, 12 and 24 hours depicted in fig. 3a, 3b, 3c and 6d respectively, show the absorbance peaks at 3500 $cm^{-1}$ and 1100 $cm^{-1}$ which are due to the O-H stretching vibration, and this can be related to the water present in the sample. For the hydrochars carbonized for 6, 12 and 24 hours indicated by figures 5b, 5c, and 5d, the peak at 1770 $cm^{-1}$ are due to the C=O stretching vibrations of the carbonyls (aldehyde, ketones esters, carboxylic acids). Another feature of the hydrochar spectra is the presence of the oxygen containing bonds (C-O) demonstrated by the peaks at 1600 $cm^{-1}$ for all three reaction times and temperatures. For the hydrochar carbonized for 6 hours at temperatures of 180, 230 and 275$^0$C in fig. 3b, the band at 700 $cm^{-1}$ corresponds to the C-Cl and C-Br stretching of the carbon-halogen bonds, while the C-H vibration in the methyl, methylene and methyne groups can be assigned a band of 1190 $cm^{-1}$. The same trend was observed for carbonization times of 12 hours and 24 hours corresponding to temperatures 180, 230 and 275$^0$C, are shown in fig. 3c and 3d. The presence of CaO-H indicated by the peak at 3000 $cm^{-1}$ at all the different temperatures (fig. 3b, 3c, and 3d) are associated to the presence of calcium carbonate and calcium phosphate in the feedstock. Similarly, the peak at 900 $cm^{-1}$ indicating the presence of Si-O was attributed to the presence of silicon dioxide in the original feedstock.

### 3.3.2 XRD Spectra of Hydrochar

The XRD pattern highlighted in fig. 4a, 4b and 4c represent the physical mineral components and crystal structures of the hydrochars at reaction times 6, 12 and 24 hours, respectively. The PDXL database search result showed patterns largely matched calcite (blue), and some peaks corresponding to n-acetyl- dl-valine (green/pink). Calcite is likely from calcium carbonate often used as an additive in medicine tablets, while n-acetyl- dl-valine will have to be confirmed through other chemical analysis.

### 3.3.3 SEM Images of Hydrochar

The hydrochars were analyzed by SEM imaging, to follow the changes in their surface morphology during the process. From fig. 5a, 5b and 5c, it is observed that the microstructure of the material appears to be irregularly shaped. It should be noted that with an HTC reaction times of 6, 12 and 24 hours, when the temperature increases from 180 $^0$C to 230 $^0$C and 275 $^0$C, the surface appearance does not change significantly which is in agreement with previous works with other biomass materials such as orange peels (Fernandez et al., 2015). Nevertheless, the hydrochar carbonized at 180 $^0$C at times 6, 12 and 24 hours appeared to be brownish-black in color compared to the hydrochars at 230 and 275 $^0$C (at times 6, 12 and 24 hours) which appeared black suggesting incomplete carbonization at 180 $^0$C.

### 3.3.4 Elemental Analysis of Hydrochars

The elemental carbon content of the hydrochar is often carefully considered because the more carbon the hydrochar possesses, the greater it's energy value. Also, more carbon in the char is preferable for the production of carbon-based materials, carbon sequestration, and soil amendment (Reza et al., 2014, Titirici et al., 2010, Zhu et al., 2014). From the carbon densification calculation, fig. 6, it is observed that the hydrothermal treatment of pharmaceuticals led to an increase in the elemental carbon content of the hydrochar from 48.1 % (FS; Feedstock) to 62.4% in the case of PH24_230 . Carbon is the most abundant element and the weight percent of carbon in the hydrochar increases as the temperature increases from 180 to 230 to 275 $^0$C. The increasing trend suggests that carbonization of the product can be achieved at higher temperatures (Kalderis et al., 2014).

### 3.3.5 Higher heating values of the hydrochars and pyrochars

The higher heating value (HHV) of the feedstock, hydrochars, and pyrochars are compared in table 3. HHV of the pharmaceuticals was 12.00 MJ/kg and decreased up to 10.35 MJ/kg for PH6_275 and 4.01 MJ/kg for PH-TGA1_700. The HHV of the hydrochars was similar, ranging from 10.35 to 12.22 MJ/kg, however, as can be seen in fig. 7, by varying reaction time different trends are shown. At 6 hours reaction time the heating value decrease with temperature almost linearly; at 12 hours, a maximum is reported at 230°C and a large decrease occurs at 275°C. In the case of 24 hours residence time the HHV is nearly constant, showing only a minor decrease at 230°C. The pyrochars possessed even lower HHV: 5.42 MJ/kg for PH-TGA1_500 and 4.01 MJ/kg for PH-TGA1_700; this trend can be attributed to the higher de-volatilization of organic matter with reaction temperature and because of the elevated concentration of inorganics in the feedstock (especially calcium), which were not de-volatilized in the pyrolysis process and behaves as "inert matter".

By considering these values, it can be said that the hydrochar from HTC of waste pharmaceuticals could be potentially used as a solid fuel even if its heating value is very low. However other more valuable applications should be investigated. On the contrary, slow pyrolysis led to pyrochars with extremely low HHV, because of the consistent volatilization of organic material, making them unsuitable for combustion.

### 3.3.6 BET analysis of the hydrochars and pyrochars

In table 4 the results of the BET analysis regarding the surface area, total pore volume and average pore diameter of the feedstock, hydrochars (PH24_275, PH12_275, PH6_275, PH24_230) and pyrochars (PH-TGA1_500, PH-TGA1_700) are reported. The feedstock had an extremely low surface area (0.48 $m^2/g$), a total pore volume of $1.75 \cdot 10^{-3}$ $cm^3/g$ and an average pore diameter of 14.71 nm. The hydrochars possessed a surface area that is one order of magnitude higher; the hydrochar produced at 24 hours and 230°C had the highest value (2.36 $m^2/g$). PH6_275, PH12_275, and PH24_275 had a surface area of 1.21, 1.82 and 1.77 $m^2/g$, respectively. Regarding the hydrochars produced at 275°C are concerned, both total pore volume and average pore diameter increased with residence time, but it can be seen that when the time change from 12 to 24, the growth is lower, confirming that the reaction equilibrium was reached after 12 hours. However, the highest values of the surface area, pore volume, and average pore diameter were obtained by PH24_230. As far as the pyrochars are concerned, the values of surface area were two orders of magnitude higher if compared to the feedstock, reaching 20.24 and 63.15 $m^2/g$ for PH-TGA1_500 and PH-TGA1_700, respectively. Also, the total pore volume is one order of magnitude higher, and the average pore diameter is one order of magnitude lower. By increasing the pyrolysis temperature both surface area and total pore volume increased, but at the same time, the average size of the pores decreased, changing from 4.74 to 2.95 nm. The different adsorption behavior of the samples was confirmed by the isotherms of nitrogen adsorption, shown in fig. 8a. The feedstock and the hydrochars show relatively similar low adsorption capacity if compared with the pyrochars curves. These latter curves are quite unusual because the adsorption and the desorption do not close by reducing the pressure ratio. On the contrary, the isotherms of the feedstock and the hydrochars, reported in fig. 8b, can be classified as type IV.

### 3.3.7 Batch Adsorption Experiments

Experiments were conducted to analyze the effect of contact time on adsorption of the $Pb^{2+}$ ions by the hydrochars. The results from the hydrochars at different temperatures and reaction times were compared to the analytical reagent activated carbon. As shown in fig. 9a, 9b and 9c, the rate of adsorption is initially fast within the first 10 minutes, and equilibrium is gradually attained. This can be explained by the availability of adsorption sites within the first minutes which are later saturated with the contaminant. At the start of the adsorption process, there is a high concentration gradient between the adsorbate concentration in solution and the adsorbents' surface; as the contaminant loading on to the adsorbent increases, the gradient reduces hence slower uptake (Oladoja et al., 2008).

From fig. 9a, it is shown that the hydrochar PH6_230 competes closely with the AR-AC with almost a 96% for the PH6_230 versus 95% removal for the AR-AC after 60 minutes. For the 6 hour reaction time, the next best adsorbent is the PH6_275 followed by the PH6_180 obtaining a 93% and 84% removal, respectively. It should be noted from fig. 9a, b and c that, all the adsorbents reached equilibrium after 10 minutes except for PH_t_275 (where t= 6, 12, 24 hours) which reached equilibrium after 40 minutes (fig. 9a), 20 minutes (fig. 9b) and 40 minutes (fig. 10c). When comparing the adsorbents for the 12-hour time period shown in fig. 9b, the following order is observed: AR-AC > PH12_180 > PH12_275 > PH12_230 with 95%, 93%, 92% and 90% removal, respectively. From fig. 9c, the PH24_230 outperformed the AR-AC, with 97% removal versus 95% for the AR-AC after 60 minutes. This was then followed by the PH24_275 which obtained 92% removal and then PH24_180 with 90% removal after 60 minutes. Table 5 summarizes the hydrochar's ability to remove $Pb^{2+}$ in comparison to AR-AC. Overall, the hydrochars ability to remove $Pb^{2+}$ ions is observed in the following order: PH24_230 > AR-AC > PH6_230 > PH6_275 > PH12_180 > PH24_275 > PH12_275 > PH24_180 > PH12_230 > PH6_180.

### 3.3.8 Kinetic Study: Pseudo Second-Order Model

A pseudo second-order model (McKay et al., 1999) was also applied to assess the kinetics of adsorption of lead on the hydrochars. The differential equation for this reaction is given as:

$$\frac{d_q}{d_t} = k_2'(q_e - q)^2 \qquad (3)$$

Integrating the equation above for the boundary conditions t = 0 to >0 and q = 0 to >0 and rearranging it gives the linearized form of pseudo second-order rate kinetics, which is shown as follows:

$$\frac{t}{q_t} = \frac{1}{k_2' q_e^2} + \frac{1}{q_e} t \qquad (4)$$

Where $q_t$ and $q_e$ are the amount of metal ion adsorbed (mg/g) at any time t and equilibrium, $k_2$ (g/mg/min) the pseudo second-order rate constant and t is the time (min). The kinetic plots between t/q versus time were plotted for the different times (6, 12, 24 hours) and the different temperatures (180, 230, 275 $^0$C) as shown in fig. 10a, b, and c. The pseudo-second order suggests that the predominant process is chemisorption whereby electrons are shared between the adsorbate and the surface of the adsorbent. Chemisorption is always limited to just one layer of molecules on the surface; however, it may be subsequent additional layers of physically adsorbed molecules (Randelovid et al., 2011; Ho, 2006). The rate constant and adsorption capacity at equilibrium were obtained from the pseudo-second order model graph.

The second-order rate constants were used to calculate the initial sorption rate, given by

$$h = k_2 q_e^2 \qquad (5)$$

The second-order rate constants and the calculated initial sorption rate values are shown in table 6. From table 6, it was observed that the regression coefficient ($R^2$) is above 0.9 showing that the model can be applied for the entire adsorption process and confirms the chemisorption of $Pb^{2+}$ onto the hydrochars.

### 3.3.9 Statistical Analysis

A simple two-factor ANOVA with replication analysis was performed to determine the interaction between the type of adsorbent and time (60 minutes) per the amount of metal ion adsorbed. The removal efficiency in terms of mg/L of $Pb^{2+}$ ions remaining after a period of time, t (t= 10, 20, 30, 40, 50 and 60 minutes) for all of the adsorbents (PH24_180, PH24_230, PH24_275, PH12_180, PH12_230, PH12_275, PH6_180, PH6_230, and PH6_275) were compared. Using the following null and alternative hypothesis: $H_0$: (amount of metal adsorbed does not depend on the time t), $H_1$: (amount of metal adsorbed does not depend on the adsorbent) and an alpha value - $\alpha = 0.05$.

The results show that the P- value for the interaction between adsorption time and type of adsorbents is observed to be significantly low (2.15E-31) and less than 0.05. This means that the interaction between the time and the type of adsorbent changes the amount of $Pb^{2+}$ adsorbed. Therefore, we reject $H_o$ and $H_1$.

The adsorbents with the best adsorption capacity (PH24_230) was compared to AR-AC using the two-factor ANOVA with replication analysis based on the following hypothesis:

$H_0$: (amount of metal adsorbed does not depend on the time t); $H_1$: (amount of metal adsorbed does not depend on the adsorbent) and an alpha value - $\alpha = 0.05$.

The P-value for the interaction was calculated to be 0.036 which is less than 0.05 showing that the adsorbents are different from each other. Consequently, the interaction between the time of the experiment and the type of adsorbent changed the amount of metal ion adsorbed.

## 4. Conclusions

In this work, waste pharmaceuticals were treated through hydrothermal carbonization to obtain hydrochars. A safe and simple to method, consisting of stainless steel reactors and a conventional oven as a heat source was used. Nine different hydrochars were obtained per experimental conditions: temperatures of 180, 230 and 275 $^0$C and reaction times of 6, 12 and 24 hours. It was observed that longer carbonization times increased the hydrochars fixed carbon content whereas the oxygen content and  the hydrochar yield decreased. Although not as competitive as the reagent grade activated carbon, the resulting hydrochars proved to be effective adsorbents for the removal of $Pb^{2+}$ ions from simulated wastewater with removal rates above 90% removal after 60 minutes. The kinetic studies showed that the adsorption of $Pb^{2+}$ follows a pseudo-second order kinetic model for batch study; an adsorption capacity of 4.84 mg/g was obtained for the $Pb^{2+}$ removal using the PH24_230 hydrochar. The mean level removal efficiency of the hydrochars was compared using the two-factor ANOVA with replication analysis. P-values were found to be significantly lower ($\alpha \leq 0.05$) than the amount of metal adsorbed depends on the interaction between the type of adsorbent and the adsorption time. In addition, the feedstock was converted to pyrochar by means of slow pyrolysis and its characteristics, in terms of higher heating value and BET analysis, were compared with the hydrochars produced at 275°C and with the one obtained at 24h and 230°C. The HHV of the hydrochars were similar and the one produced at 12 hours and 230°C possessed the highest value (12.22 MJ/kg), which was slightly higher than the feedstock's. Slow pyrolysis led to pyrochars with a nearly halved heating value because of a greater volatilization of organic matter. By considering these values, it can be said that the hydrochar from HTC of waste pharmaceuticals could be potentially used as a solid fuel even if its heating value is very low. However, other more valuable applications should be investigated. On the contrary, slow pyrolysis led to pyrochars with even lower HHV, because of the consistent volatilization of organic material, making them unsuitable for combustion. The BET analysis showed that the surface area of the pyrochars (highest value 63.15 $m^2/g$) was one order of magnitude higher than the one of the hydrochars (highest value 2.36 $m^2/g$). Notwithstanding these very low values of the specific surface area, the hydrochars proved to be very effective adsorbents for $Pb^{2+}$ ions. This work demonstrated that the HTC process offers an attractive and alternative technique for the conversion of waste pharmaceuticals to value-added products. Further work is required to gain a better understanding of the underlying process and to characterize the resulting liquid and gas phase products and their related properties and to identify applications for these products.

## 5. Acknowledgments

This study is based upon work supported by the United States the National Science Foundation (NSF, under Grant No. HRD 1436222). Any opinions, findings, and conclusions or recommendations expressed in this material are those of the author(s) and do not necessarily reflect the views of the NSF.

## Figures and Tables

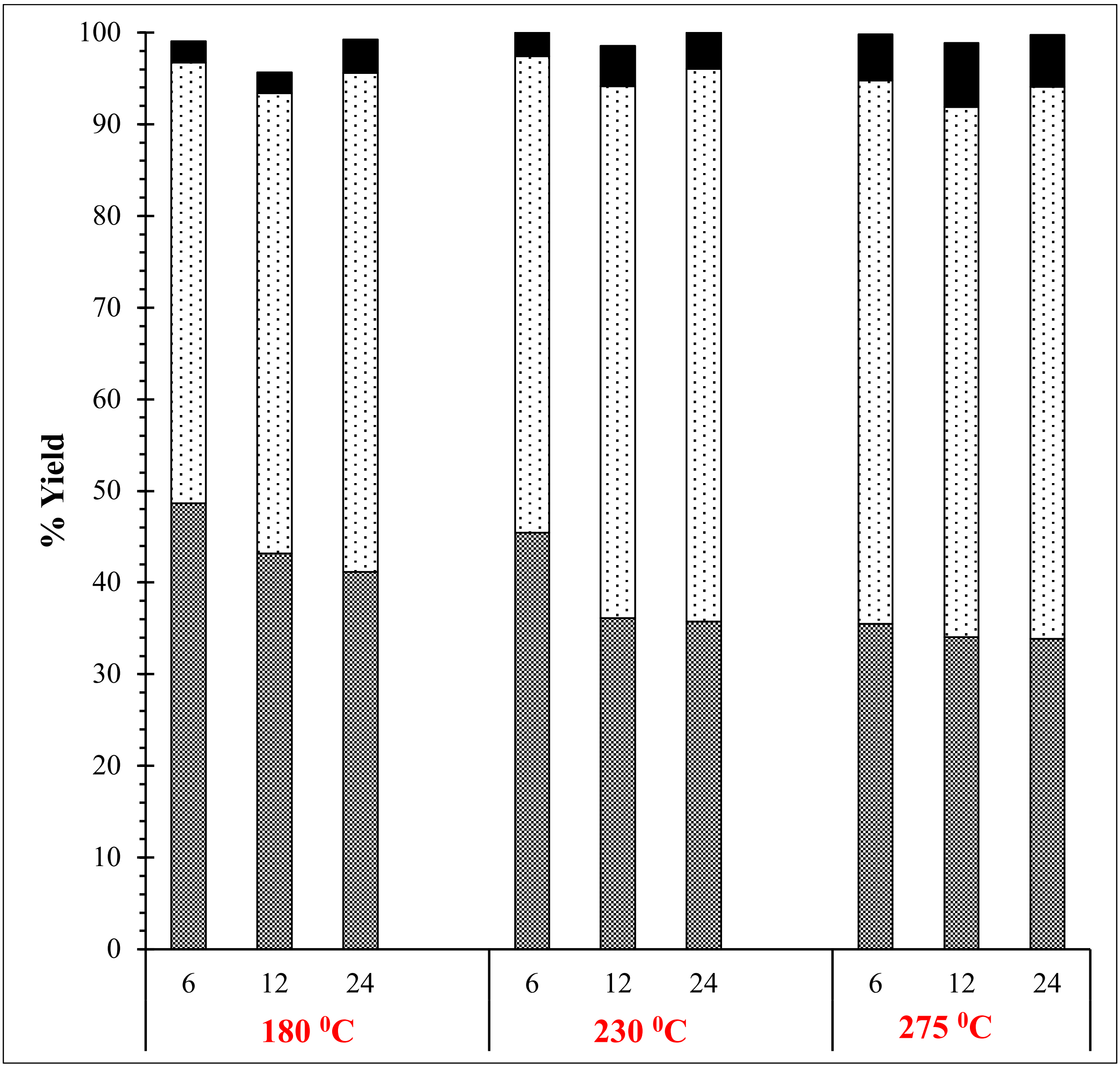

**Fig. 1.** Percentage Yield of ▨ Hydrochar, □ Liquid, ■ Gas for three HTC times 6, 12 and 24 hours

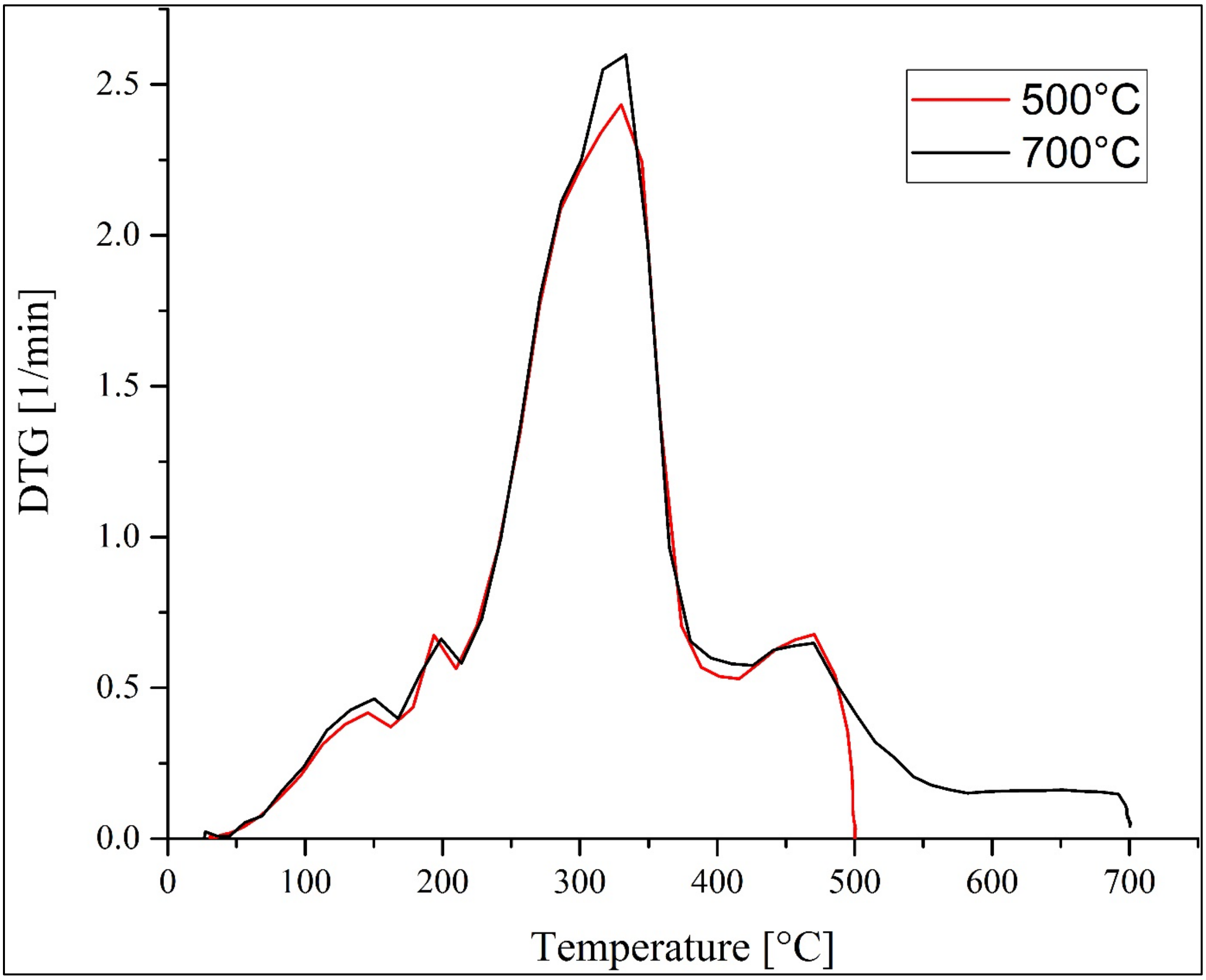

**Fig. 2.** A derivative of the TGA during pyrochar production versus time.

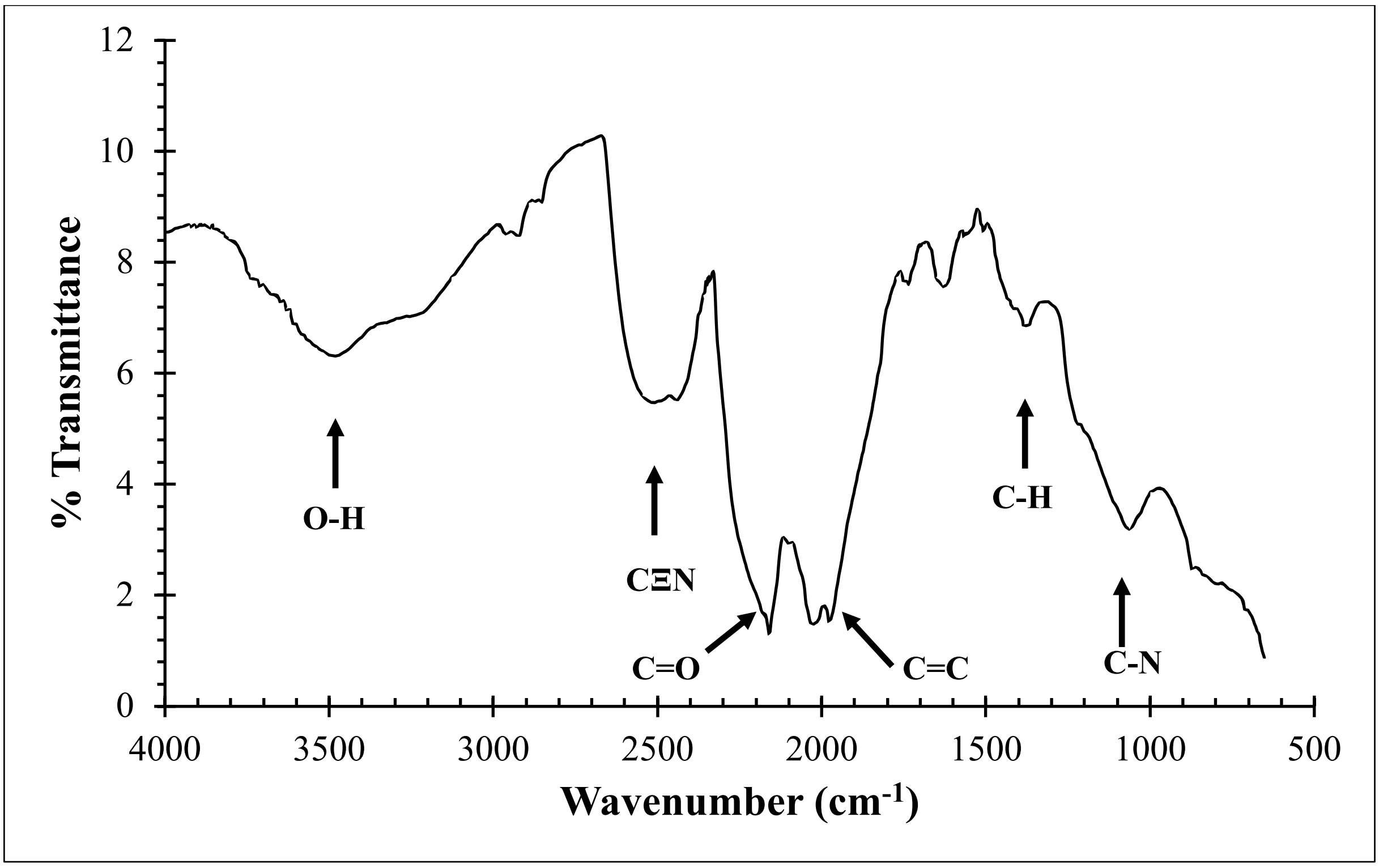

**Fig. 3a*:*** FTIR Spectra of Feedstock.

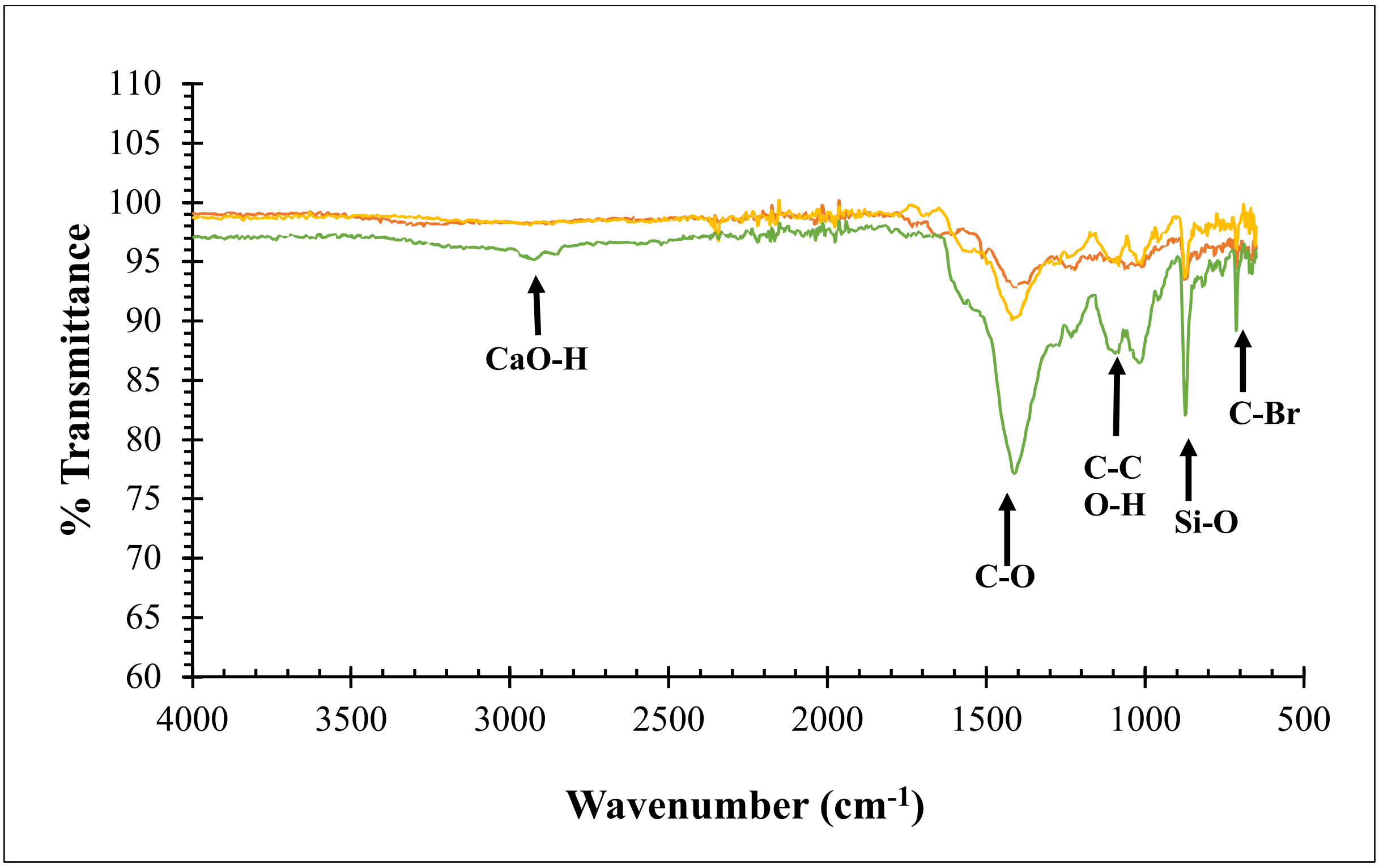

**Fig. 3b.** FTIR Spectra of the hydrochar products at HTC time of 6 hours and temperatures 180 $^0$C, 230 $^0$C and 275 $^0$C.

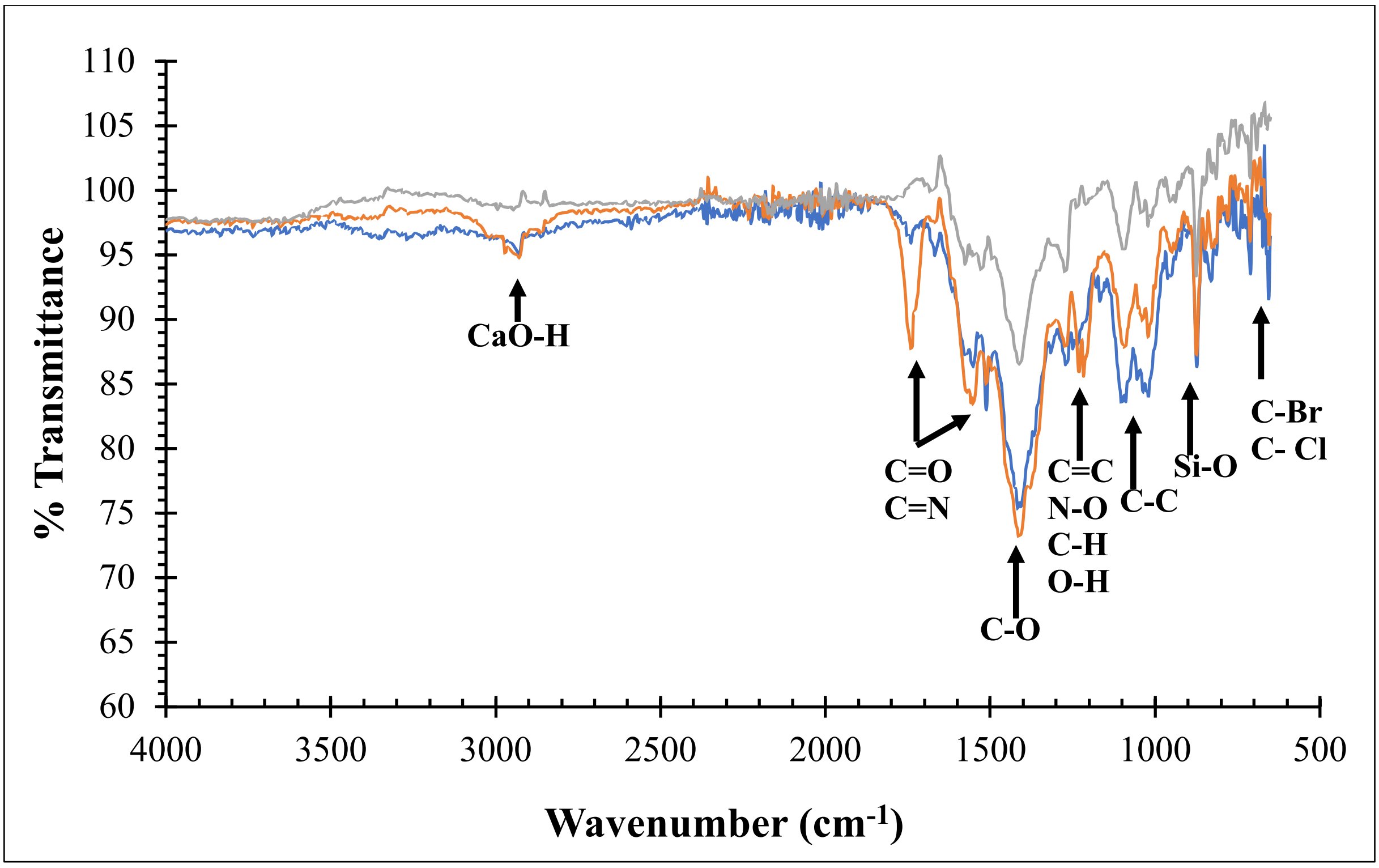

**Fig. 3c.** FTIR Spectra of the hydrochar products at HTC time of 12 hours and temperatures 180 ⁰C, 230 ⁰C and 275 ⁰C.

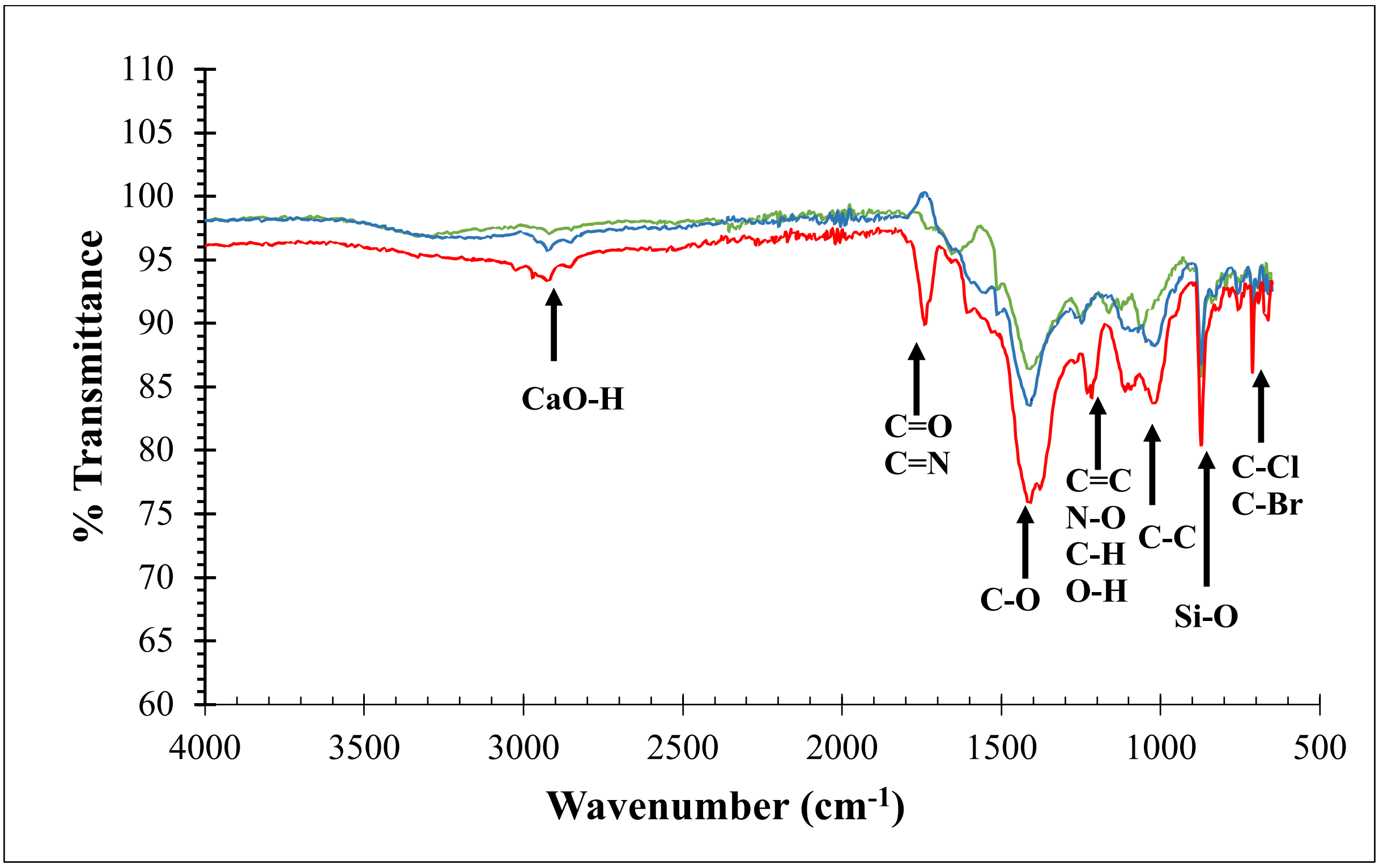

**Fig. 3d.** FTIR Spectra of the hydrochar products at HTC time of 24 hours and temperatures 180 ⁰C 230 ⁰C and 275 ⁰C.

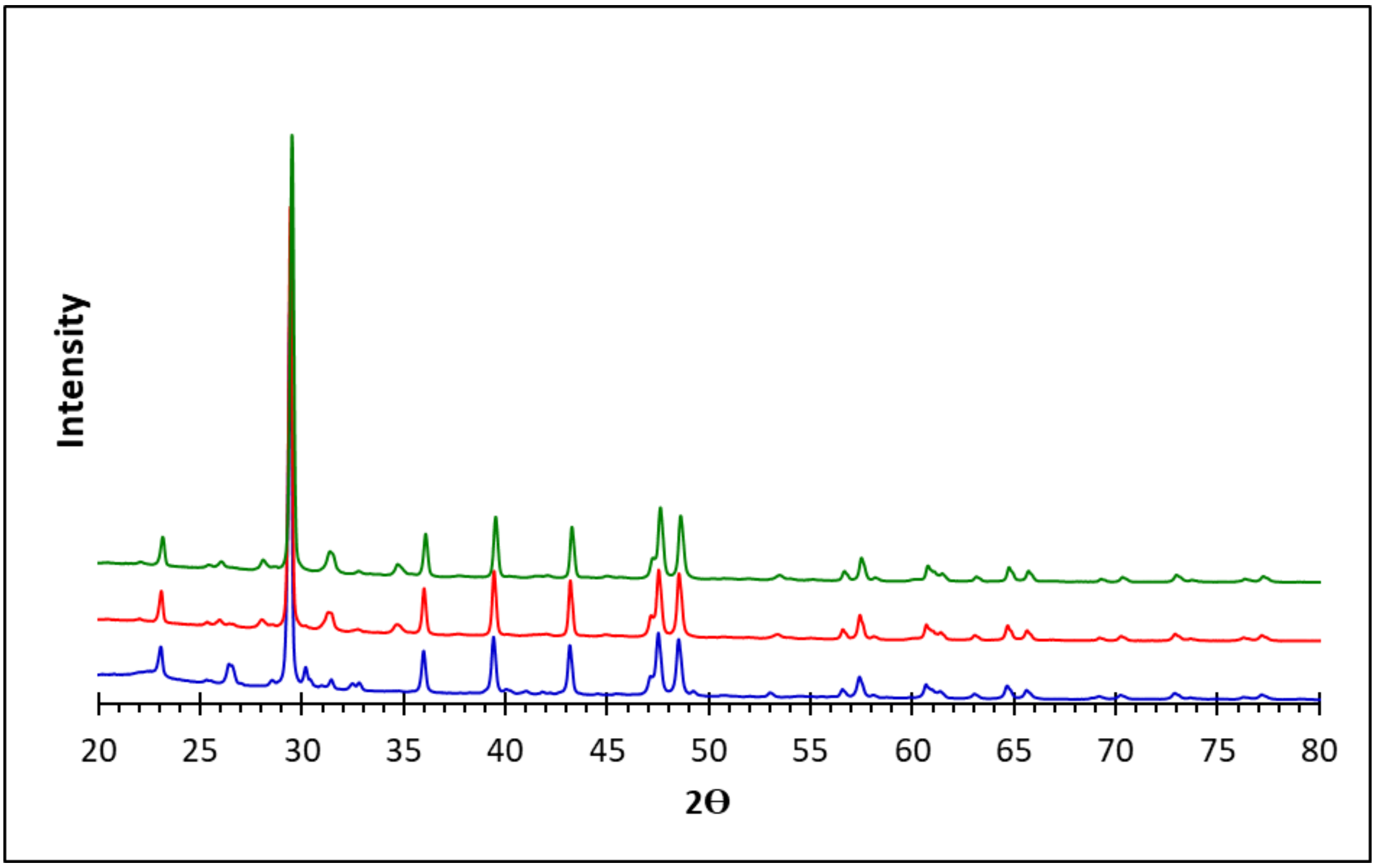

**Fig. 4a.** XRD Profiles of carbonized pharma materials at reaction time 6h and temperatures
180 $^0$C  230 $^0$C and  275 $^0$C.

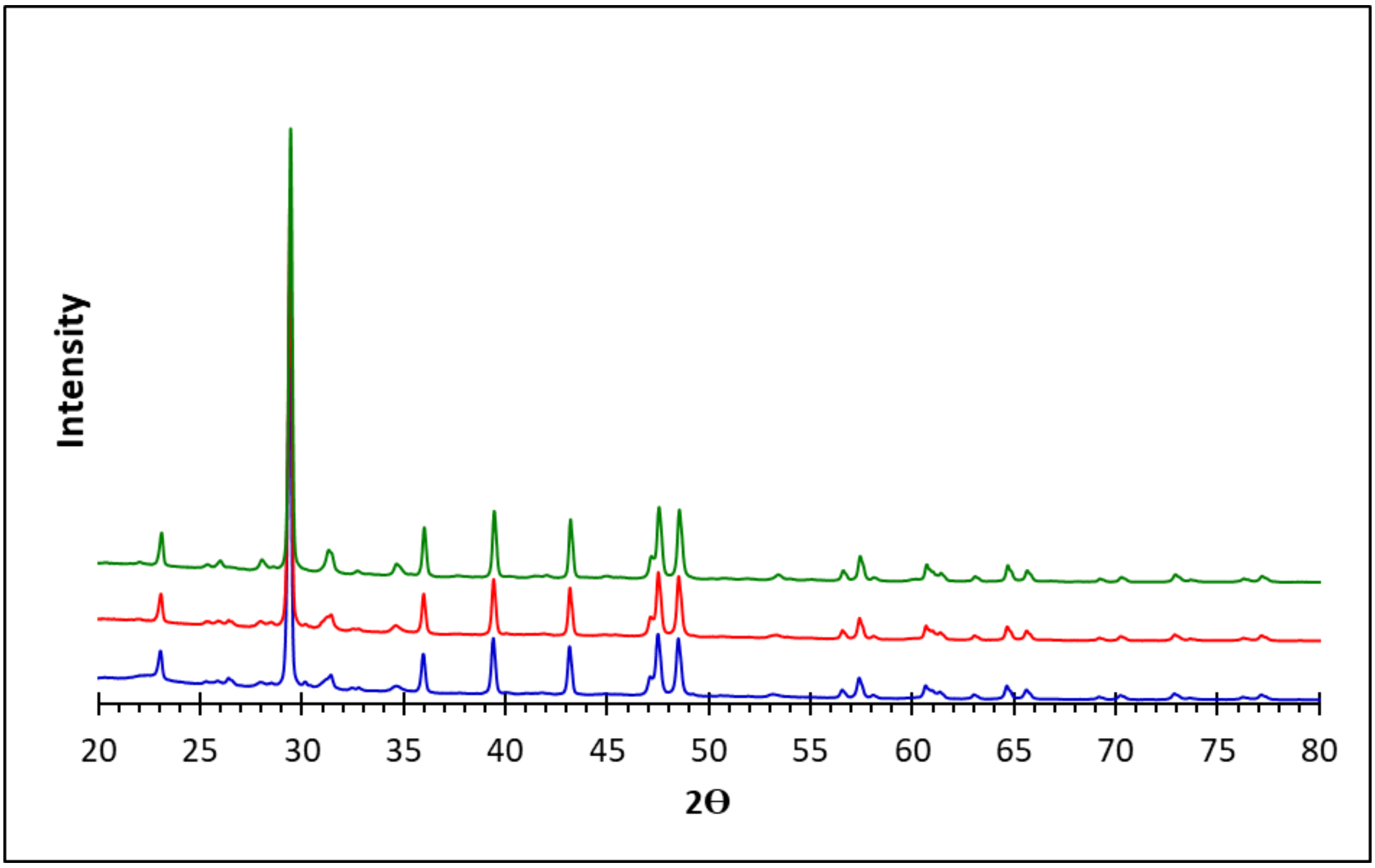

**Fig. 4b.** XRD Profiles of carbonized pharma materials at reaction time 12h and temperatures 180 $^{0}$C 230 $^{0}$C and 275 $^{0}$C.

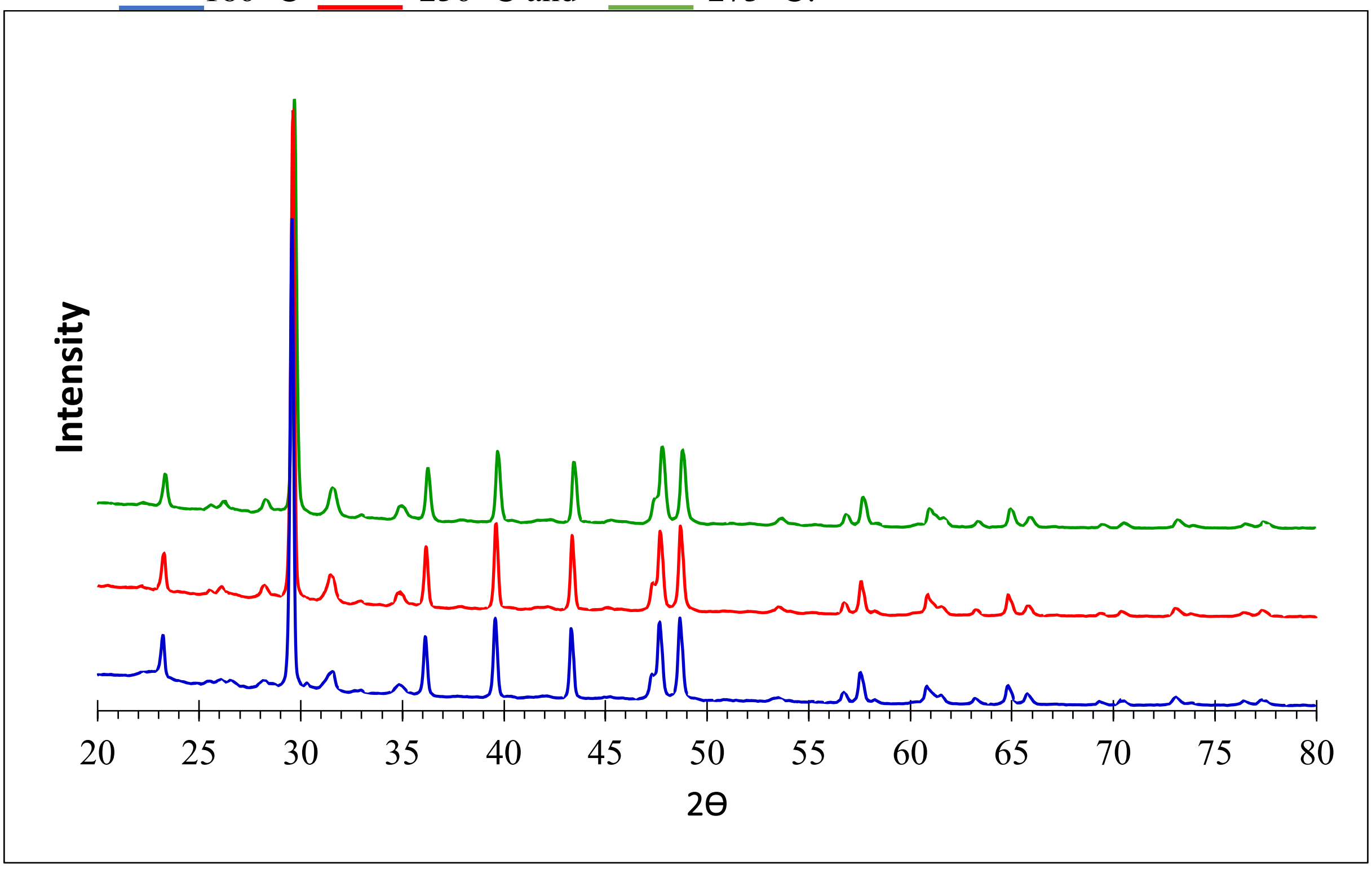

**Fig. 4c.** XRD Profiles of carbonized pharma materials at reaction time 24h and temperatures 180 $^{0}$C 230 $^{0}$C and 275 $^{0}$C.

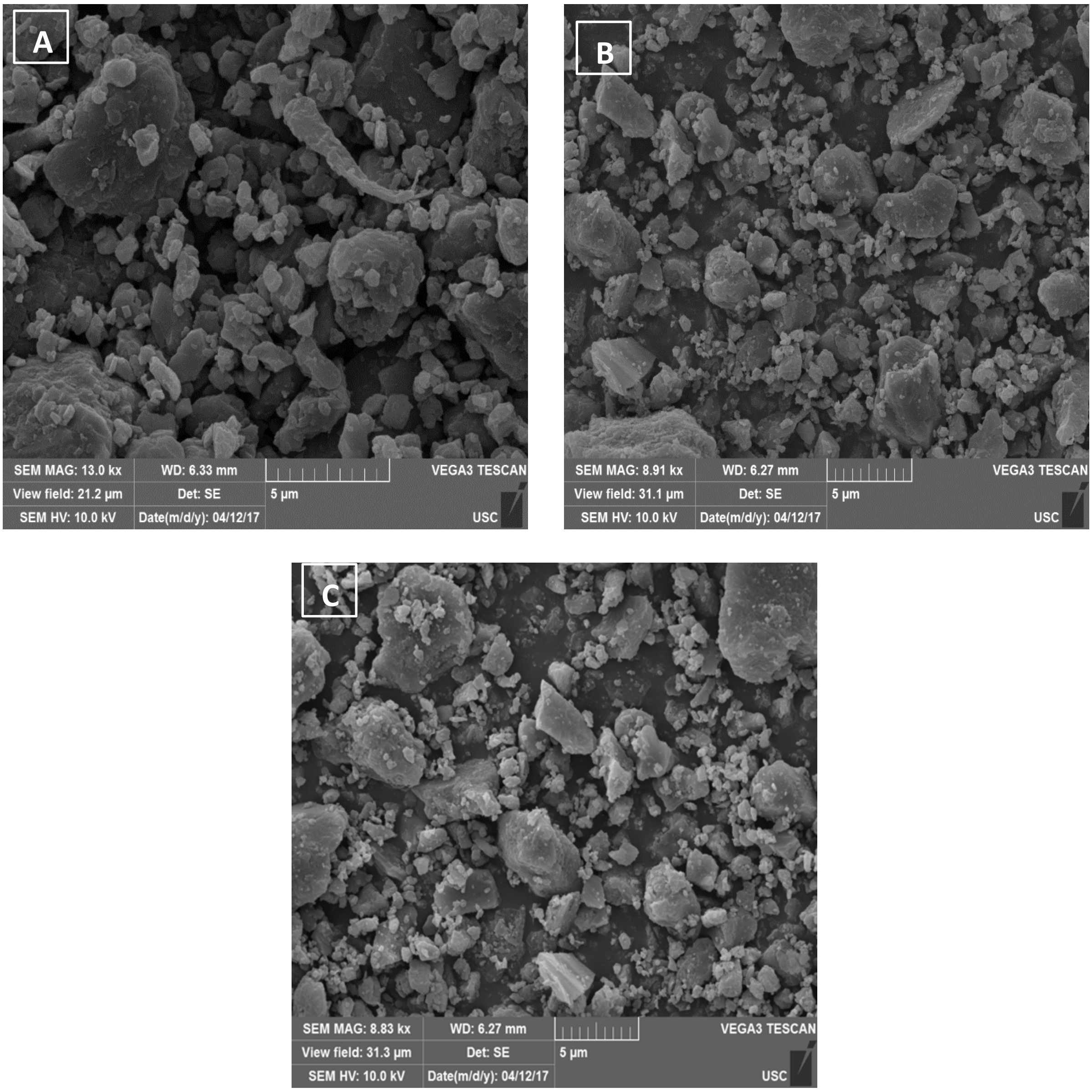

**Fig. 5a.** SEM Micrographs depicting the morphology of the hydrochars obtained after 6 hours of carbonization: (a) PH6_180 (b) PH6_230, (c) PH6_275.

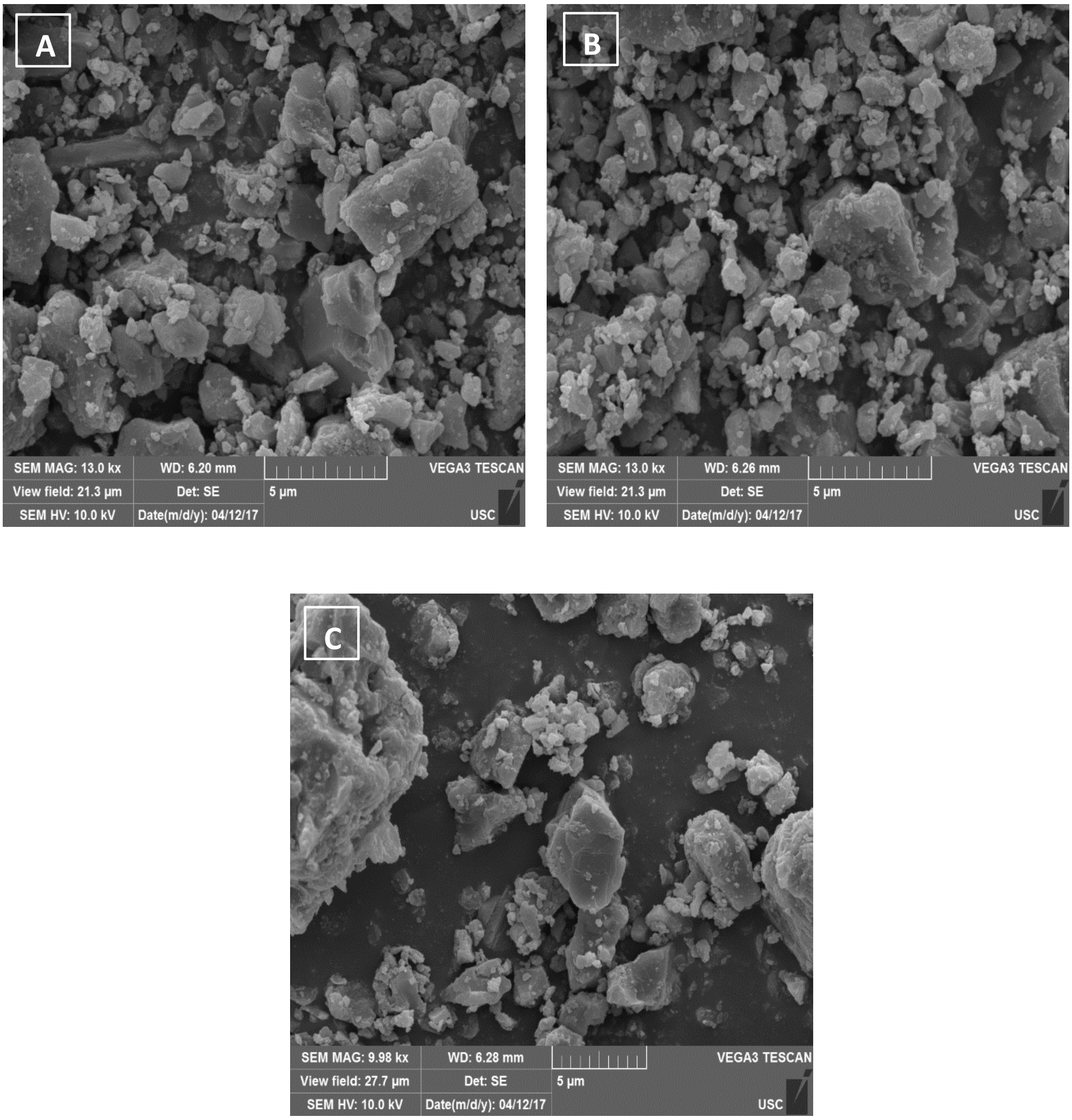

(c)

**Fig. 5b.** SEM Micrographs depicting the morphology of the hydrochars obtained after 12 hours of carbonization: (a) PH6_180 (b) PH6_230, (c) PH6_275.

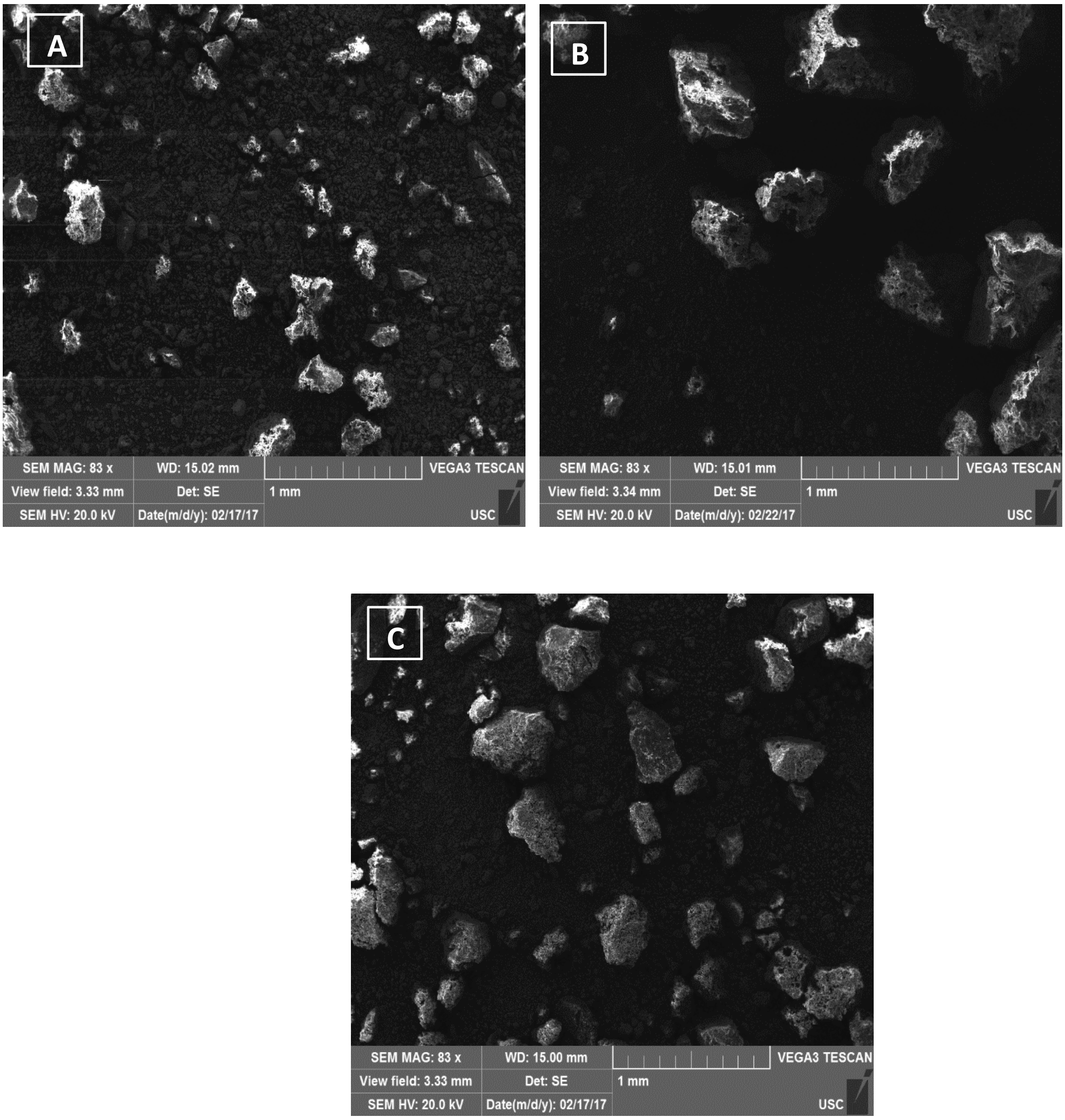

**Fig. 5c.** SEM Micrographs depicting the morphology of the hydrochars obtained after 24 hours of carbonization: (a) PH6_180 (b) PH6_230, (c) PH6_275.

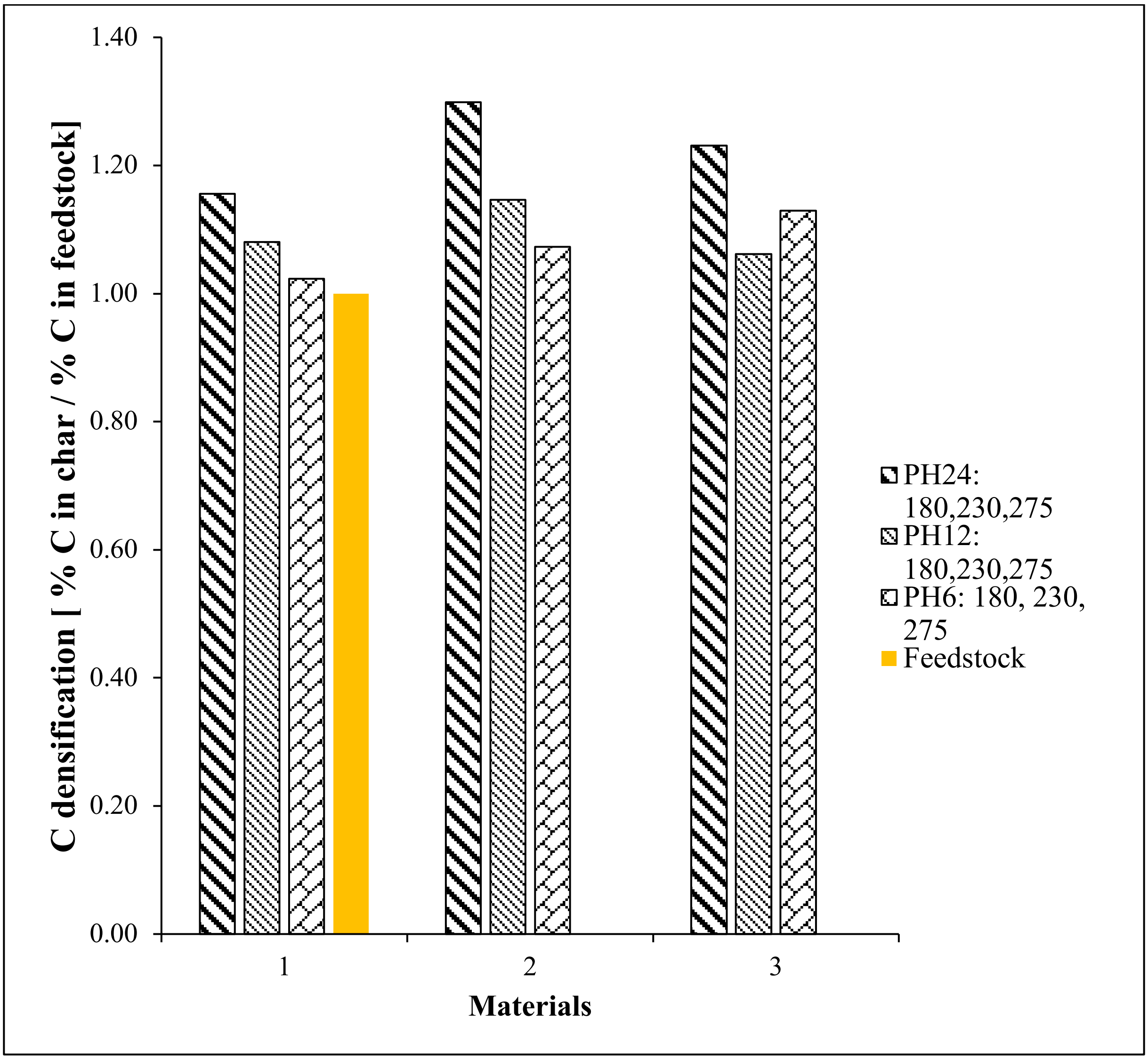

**Fig. 6.** Carbon Densification of hydrochar and feedstock

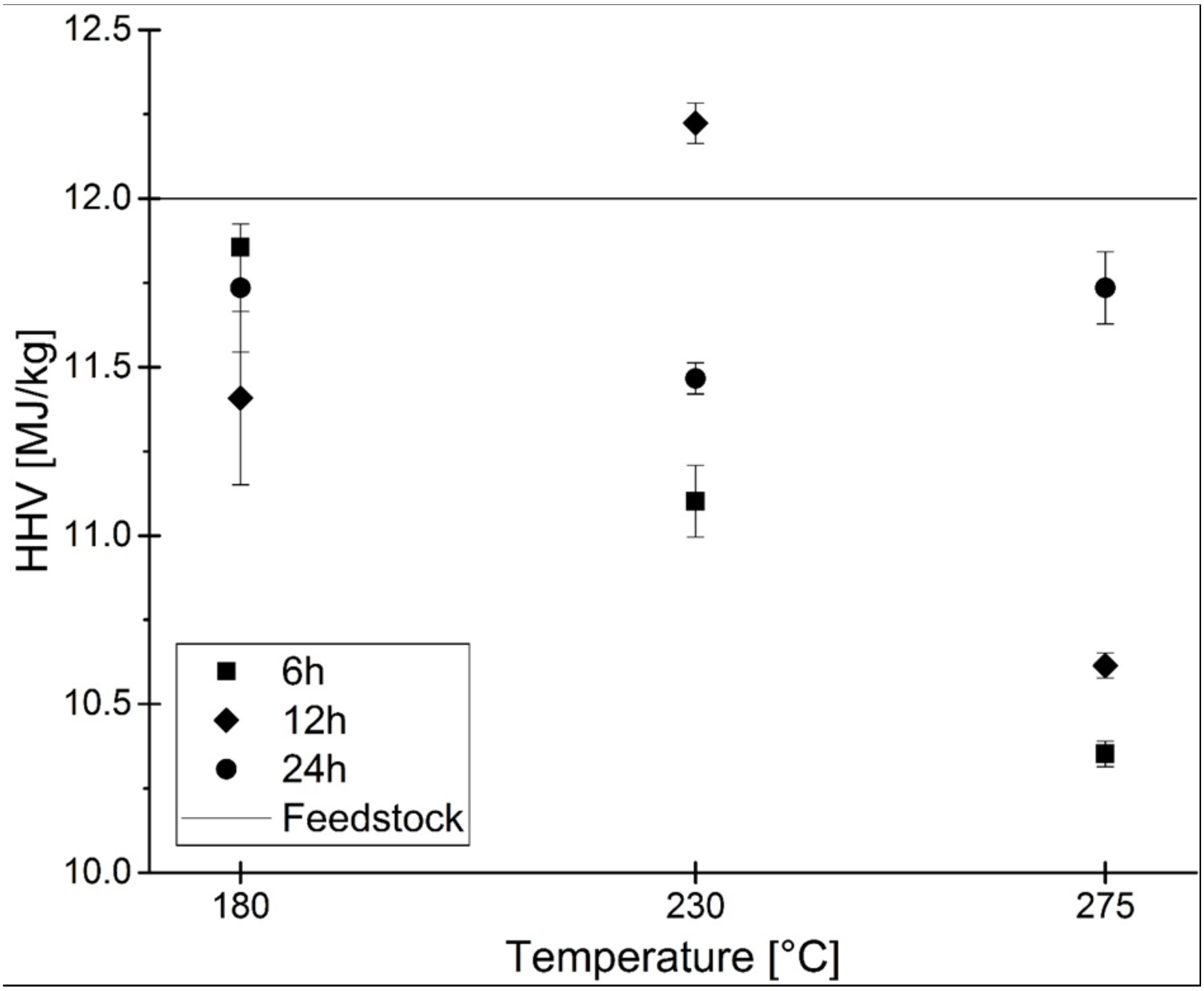

**Fig. 7.** Higher heating values of the obtained hydrochars, compared with the raw feedstock.

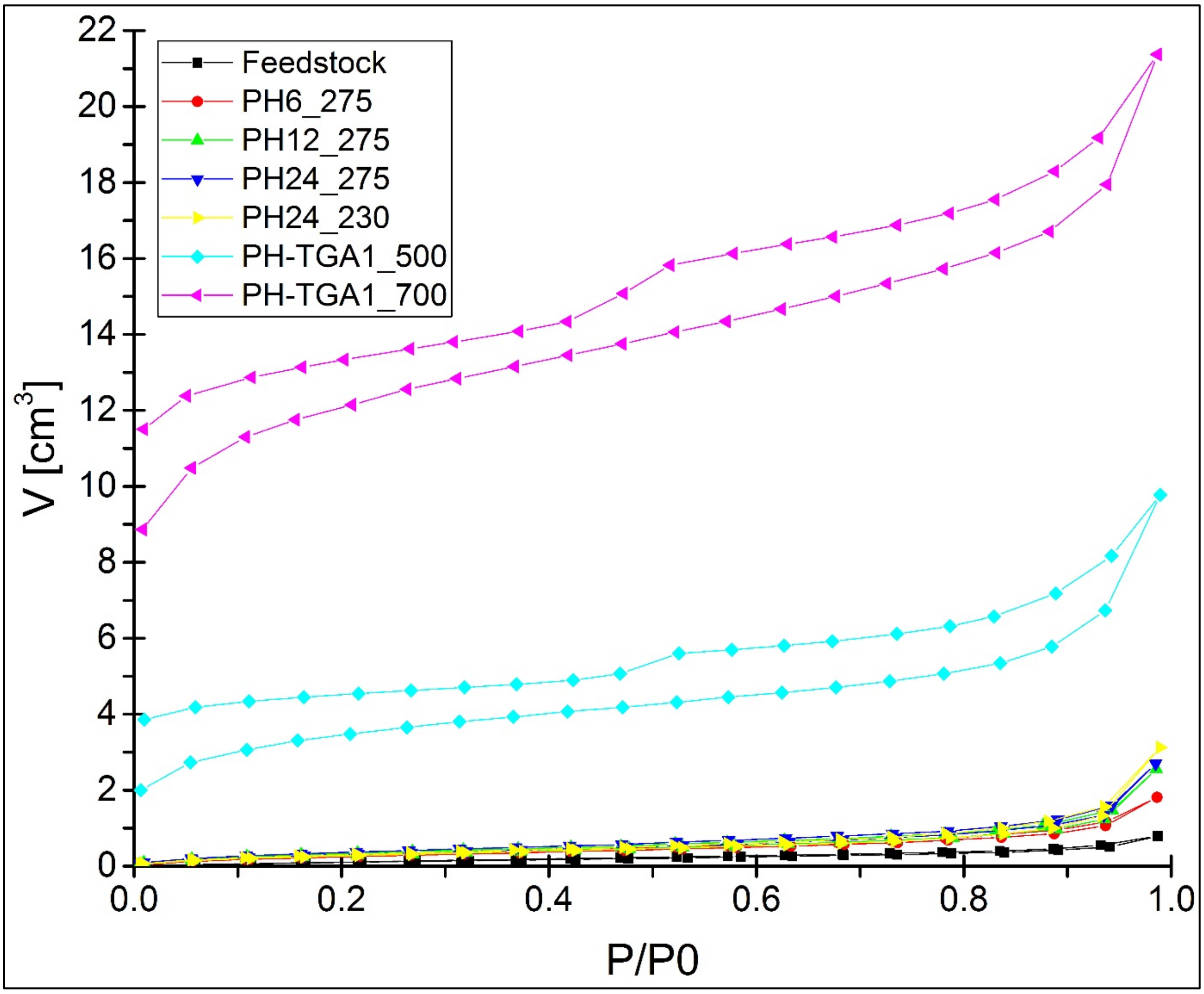

**Fig. 8a.** Nitrogen adsorption-desorption isotherms for BET analysis of feedstock, hydrochars, and pyrochars.

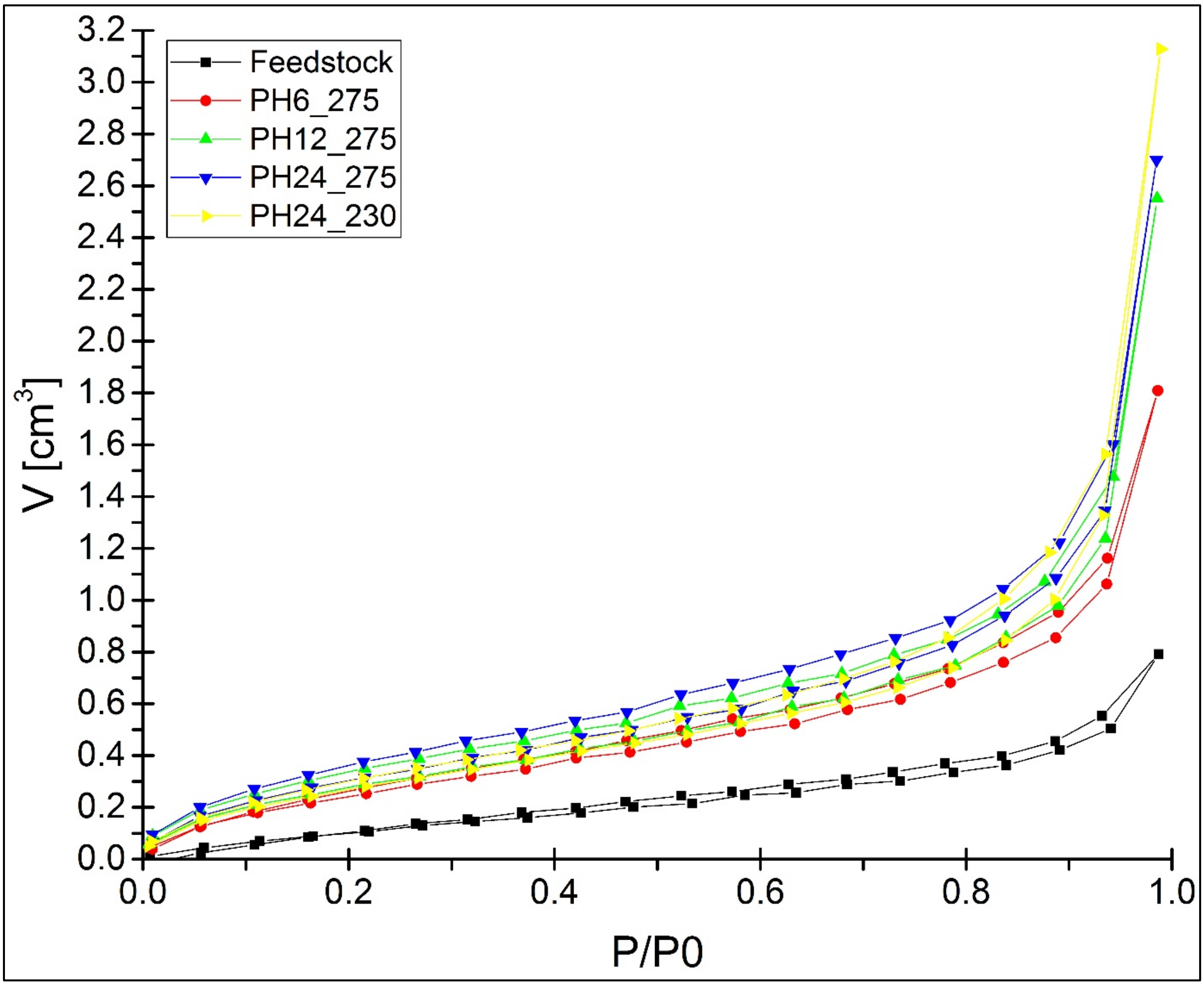

**Fig. 8b.** Nitrogen adsorption-desorption isotherms for BET analysis of feedstock and hydrochars.

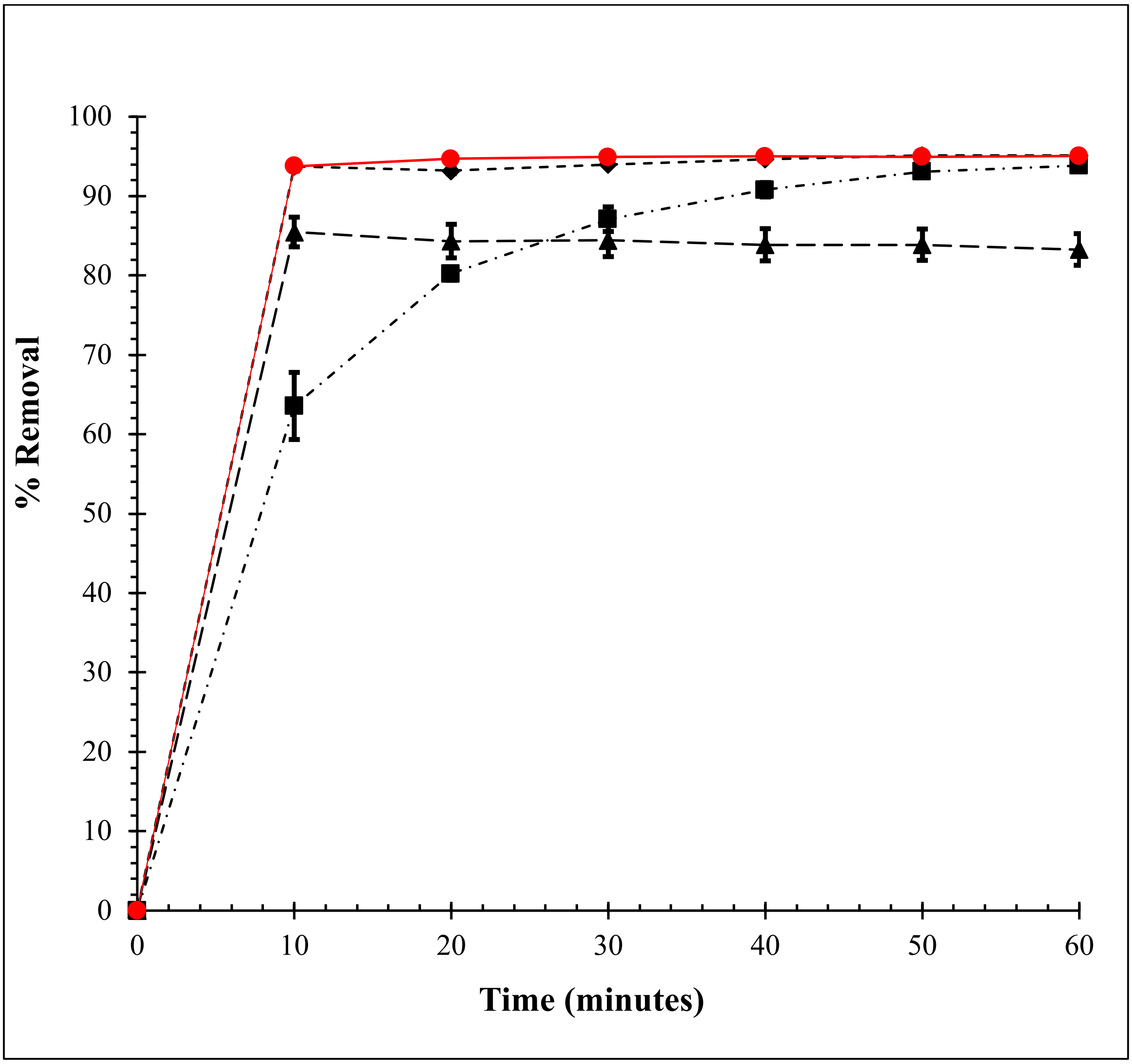

**Fig. 9a.** Batch adsorption study of lead comparing ● AR-AC to the hydrochar at HTC time 6 hours and temperatures: ▲ 180 $^{0}$C, ◆ 230 $^{0}$C, and ■ 275 $^{0}$C.

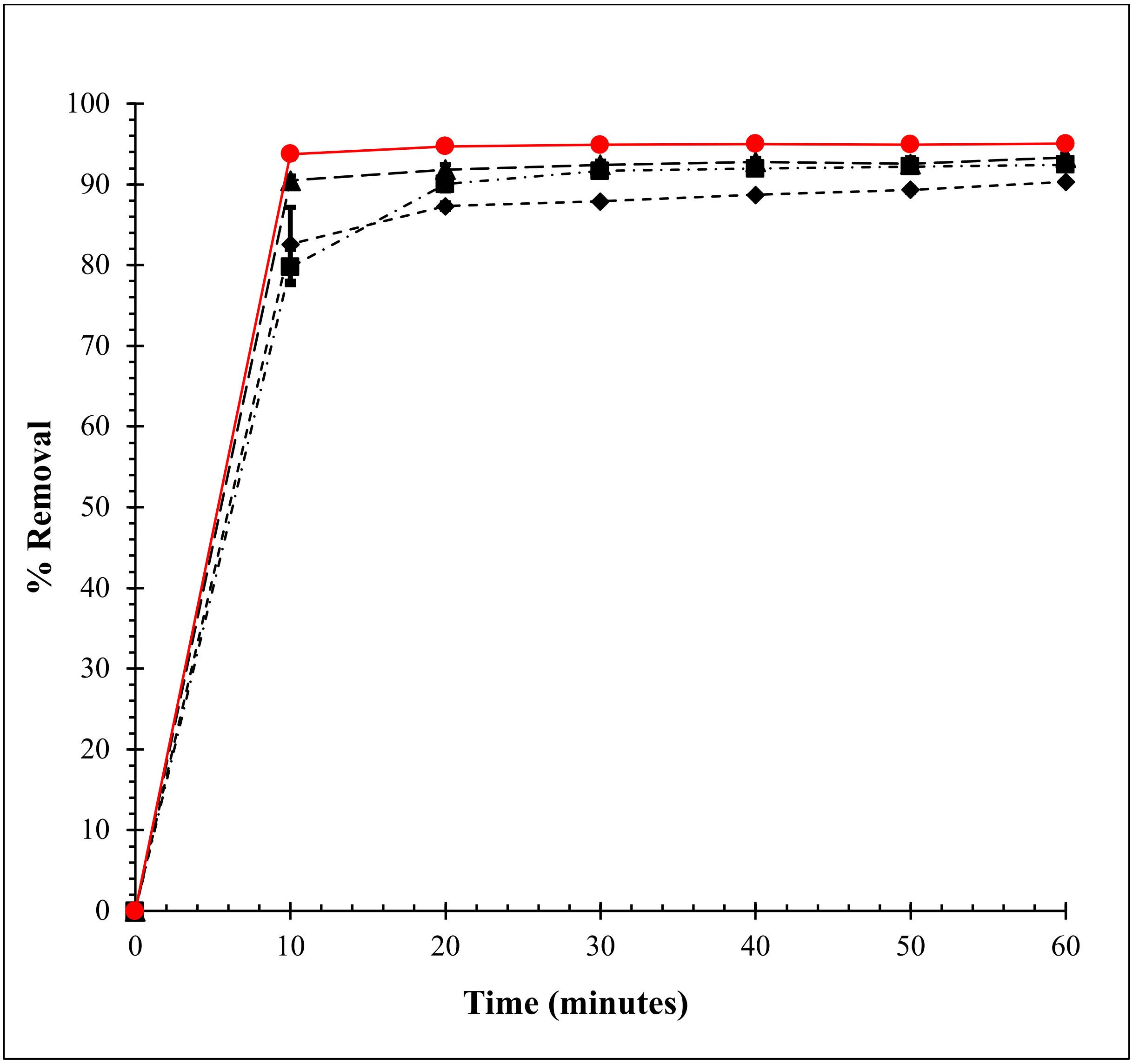

**Fig. 9b.** Batch adsorption study of lead comparing ● AR-AC to the hydrochar at HTC time 12 hours and temperatures: ▲ 180 $^0$C, ◆ 230 $^0$C, and ■ 275 $^0$C.

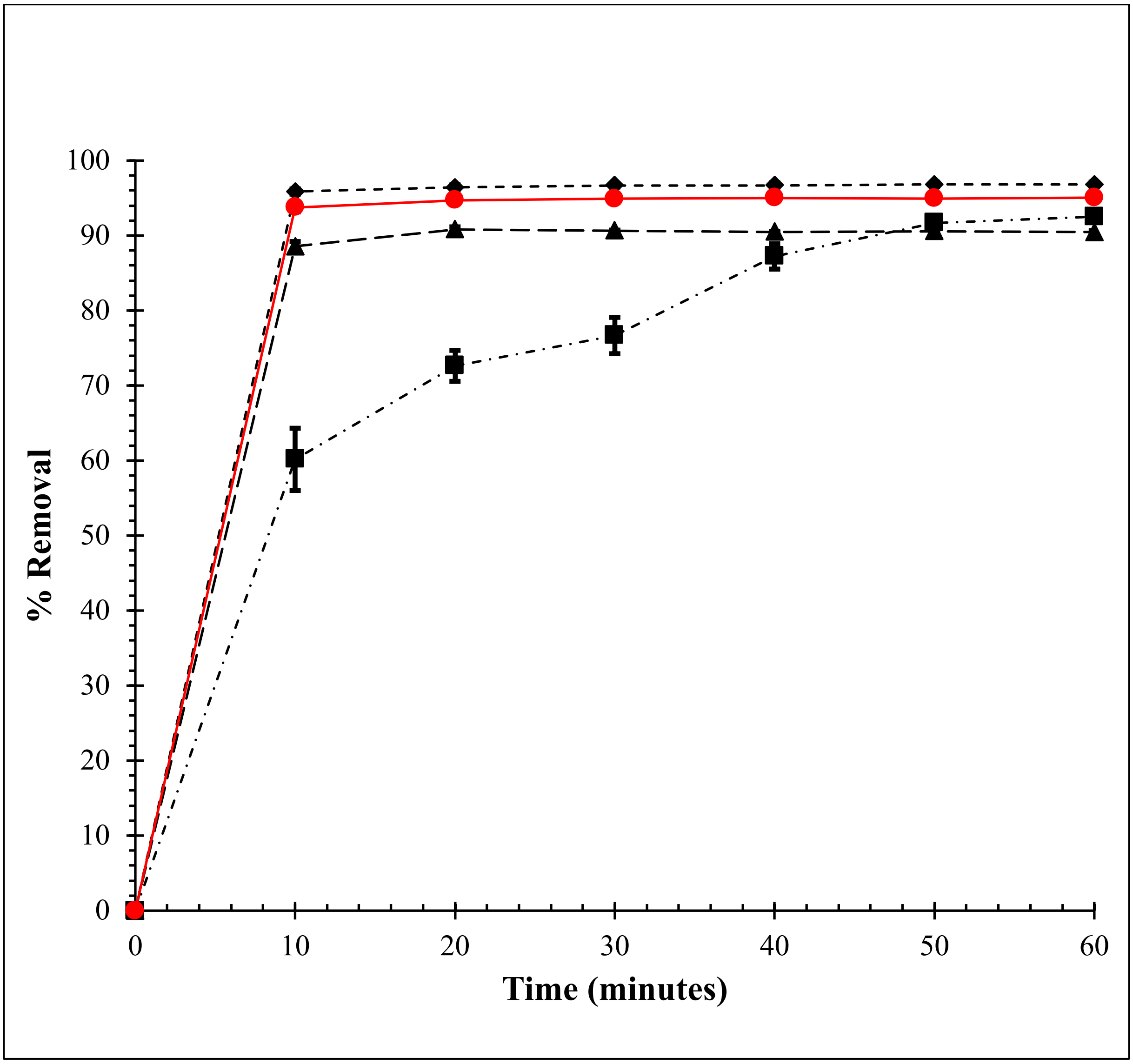

**Fig. 9c.** Batch adsorption study of lead comparing ● AR-AC to the hydrochar at HTC time 24 hours and temperatures: ▲ 180 ⁰C, ◆ 230 ⁰C, and ■ 275 ⁰C.

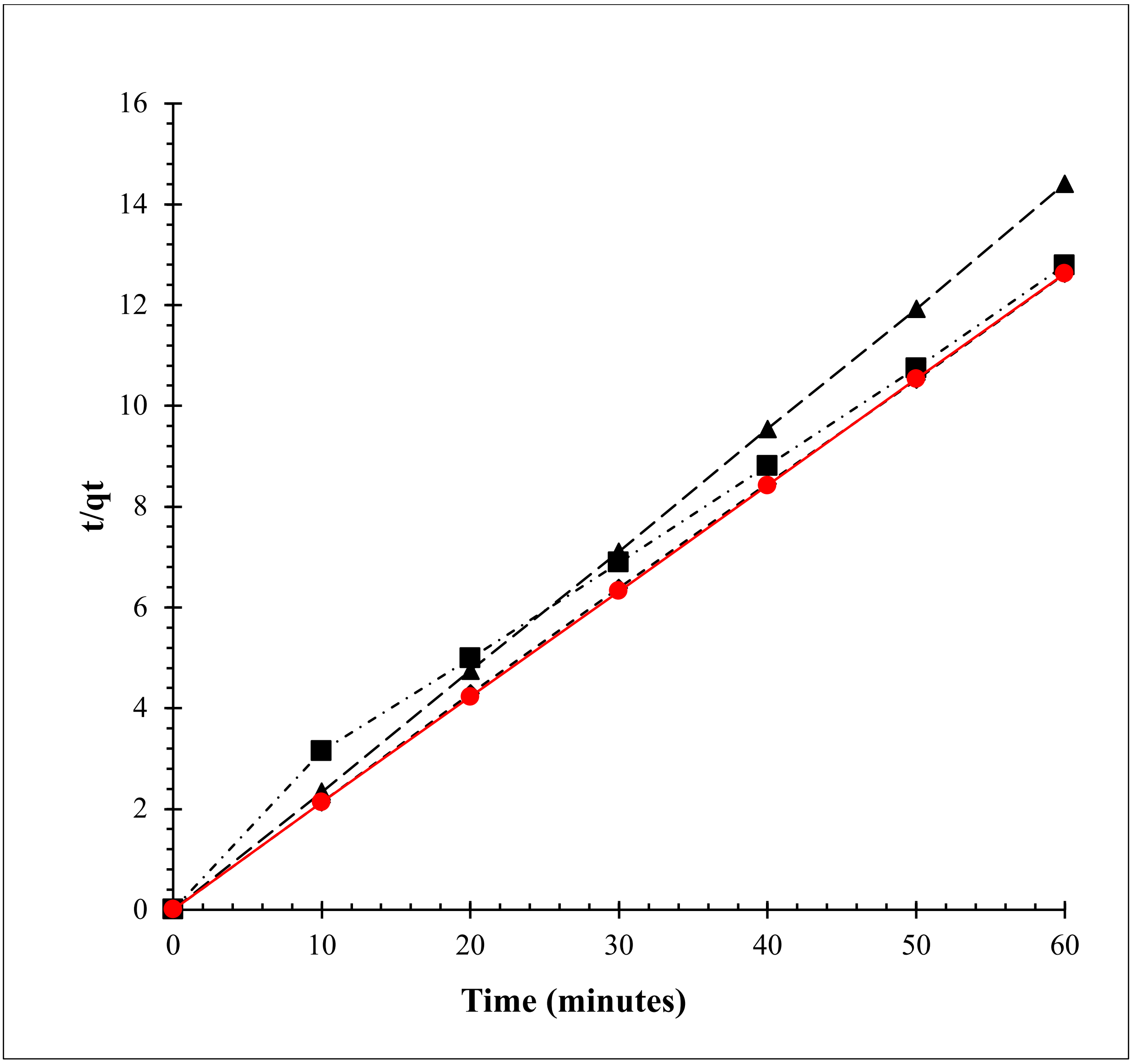

**Fig. 10a.** Pseudo second order kinetic model for $Pb^{2+}$ adsorption study comparing ● AR-AC to the hydrochar at HTC time 6 hours and temperatures: ▲ 180 $^0$C, ◆ 230 $^0$C and ■ 275 $^0$C.

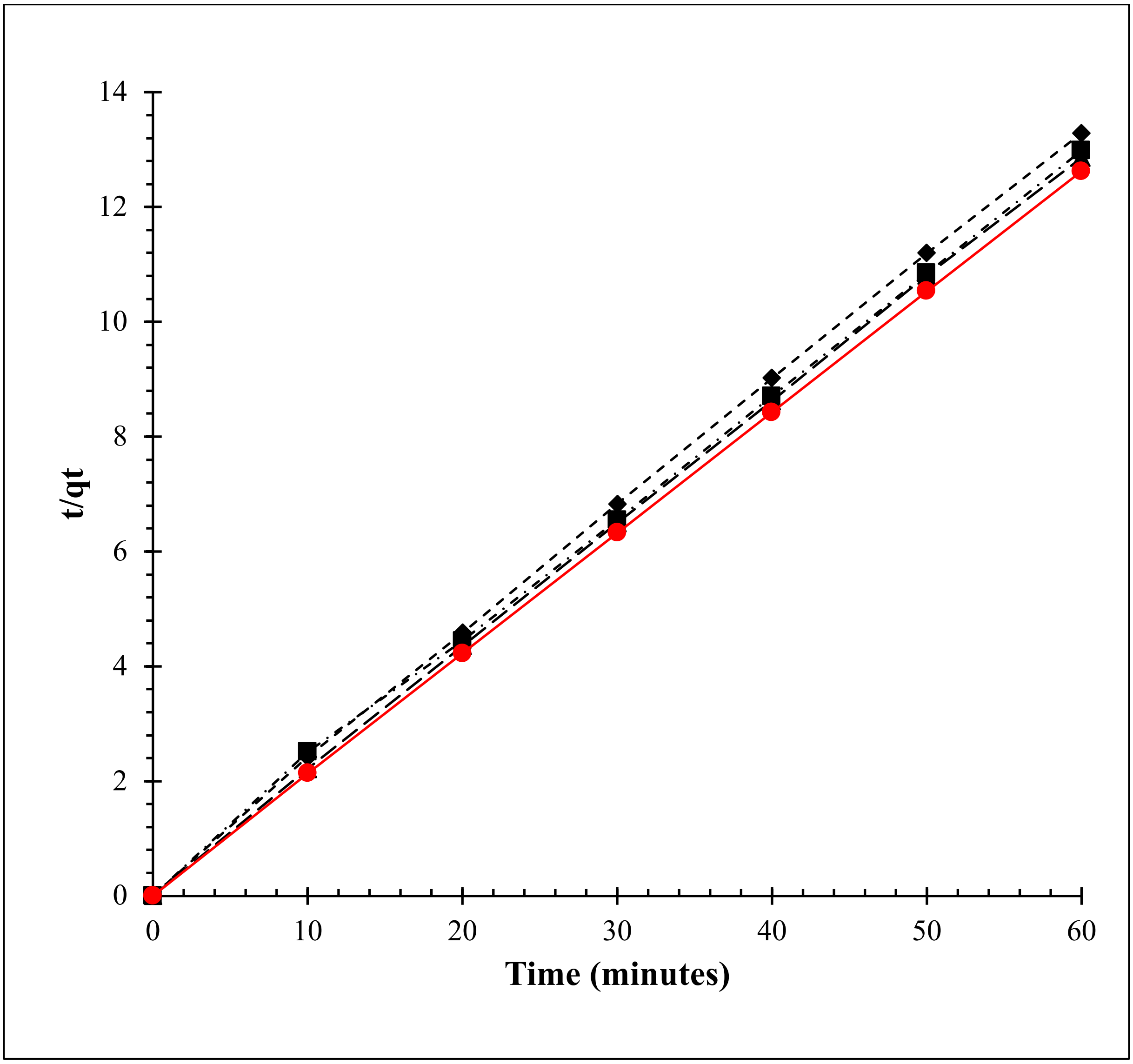

**Fig. 10b.** Pseudo second order kinetic model for $Pb^{2+}$ adsorption study comparing

● AR-AC to the hydrochar at HTC time 12 hours and temperatures: ▲ 180 $^0$C,

◆ 230 $^0$C and ■ 275 $^0$C.

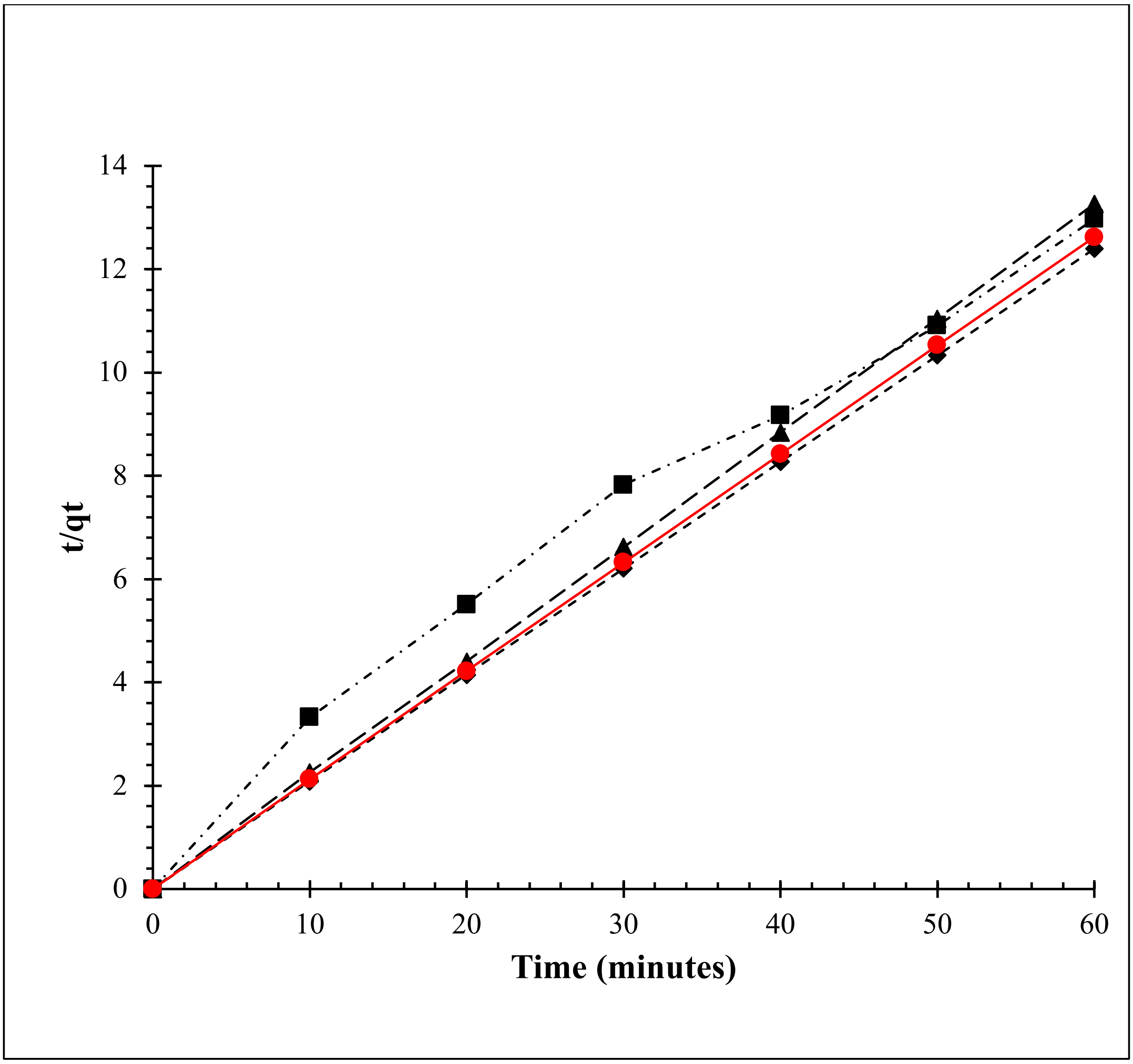

**Fig. 10c.** Pseudo second order kinetic model for $Pb^{2+}$ adsorption study comparing ● AR-AC to the hydrochar at HTC time 24 hours and temperatures: ▲ 180 $^{0}$C, ◆ 230 $^{0}$C and ■ 275 $^{0}$C.

**Table 1.** Mass balance of HTC of pharmaceuticals at three temperatures and reaction times**.**

| *Feedstock* | **HTC Temperature ($^0$C)** | **Time** | *Mass in* **Biomass (g)** | *Mass out* **Hydrochar (g)** | **Net Liquid (g)** | **Gas (g)*** |
|---|---|---|---|---|---|---|
| *Pharmaceuticals* | 180 | 6 | 40.0 ± 0.0 | 19.47 ± 0.37 | 19.21 ± 0.33 | 0.93 ± 0.04 |
| | | 12 | 40.0 ± 0.0 | 17.26 ± 0.81 | 20.09 ± 1.23 | 0.91 ± 0.34 |
| | | 24 | 40.0 ± 0.0 | 16.47 ± 0.81 | 21.76 ±0.62 | 1.46 ± 0.30 |
| | 230 | 6 | 40.0 ± 0.0 | 18.18 ± 1.05 | 20.78 ± 0.39 | 1.04 ± 0.19 |
| | | 12 | 40.0 ± 0.0 | 14.44 ± 0.16 | 23.22 ± 1.6 | 1.76 ± 0.09 |
| | | 24 | 40.0 ± 0.0 | 14.29 ± 0.22 | 24.12 ± 1.55 | 1.58 ± 0.39 |
| | 275 | 6 | 40.0 ± 0.0 | 14.20 ± 0.16 | 23.72 ± 0.45 | 2.00 ± 0.08 |
| | | 12 | 40.0 ± 0.0 | 13.62 ± 1.05 | 23.13 ± 0.98 | 2.79 ± 0.58 |
| | | 24 | 40.0 ± 0.0 | 13.55 ± 0.13 | 24.09 ± 1.42 | 2.25 ± 0.17 |

Note: ***Gas** weighs were not measured but calculated by subtracting from the solids and liquids

**Table 2.** Percentage yield of Hydrochar at three different HTC temperatures and time.

| *HTC Temperature ($^0C$)* | **Time** | *Percentage Yield %* **Hydrochar** | **Liquid** | **Gas** |
|---|---|---|---|---|
| *180* | 6 | 48.68 | 48.03 | 2.33 |
| | 12 | 43.15 | 50.23 | 2.28 |
| | 24 | 41.18 | 54.40 | 3.65 |
| *230* | 6 | 45.45 | 51.95 | 2.60 |
| | 12 | 36.10 | 58.05 | 4.40 |
| | 24 | 35.73 | 60.30 | 3.95 |
| *275* | 6 | 35.50 | 59.30 | 5.00 |
| | 12 | 34.05 | 57.83 | 6.98 |
| | 24 | 33.88 | 60.23 | 5.63 |

**Table 3.** Higher heating values of the feedstock, hydrochars, and pyrochars; the absolute standard deviation is reported in brackets.

| **Sample** | **Time (hours)** | **Temperature (°C)** | **Higher heating value (MJ/kg)** |
|---|---|---|---|
| Feedstock | - | - | 12.00 (0.40) |
| Hydrochar | 6 | 275 | 10.35 (0.04) |
| | 12 | 275 | 10.61 (0.04) |
| | 24 | 275 | 10.96 (0.11) |
| | 6 | 230 | 11.10 (0.11) |
| | 12 | 230 | 12.22 (0.06) |
| | 24 | 230 | 11.47 (0.05) |
| | 6 | 180 | 11.86 (0.02) |
| | 12 | 180 | 11.41 (0.26) |
| | 24 | 180 | 11.74 (0.19) |
| Pyrochar | 1 | 500 | 5.42 (0.16) |
| | 1 | 700 | 4.01 (0.68) |

**Table 4.** Results from the BET analysis of the feedstock, selected hydrochars and pyrochars.

| **Sample** | **Time (hours)** | **Temperature (°C)** | **Surface area ($m^2/g$)** | **Total pore volume ($cm^3/g$)** | **Average pore diameter (nm)** |
|---|---|---|---|---|---|
| Feedstock | - | - | 0.477 | $1.75 \cdot 10^{-3}$ | 14.71 |
| Hydrochar | 6 | 275 | 1.21 | $3.67 \cdot 10^{-3}$ | 12.12 |
| | 12 | 275 | 1.82 | $6.08 \cdot 10^{-3}$ | 13.36 |
| | 24 | 275 | 1.77 | $6.14 \cdot 10^{-3}$ | 13.92 |
| | 24 | 230 | 2.36 | $8.44 \cdot 10^{-3}$ | 14.29 |
| Pyrochar | 1 | 500 | 20.24 | $2.40 \cdot 10^{-2}$ | 4.74 |
| | 1 | 700 | 63.15 | $4.66 \cdot 10^{-2}$ | 2.95 |

**Table 5.** Adsorptive capabilities of obtained hydrochars and the activated carbon (control).

| Adsorbent | % Removal |
|---|---|
| PH6_180 | 83.3 |
| PH6_230 | 95.1 |
| PH6_275 | 93.9 |
| PH12_180 | 93.3 |
| PH12_230 | 90.3 |
| PH12_275 | 92.4 |
| PH24_180 | 90.5 |
| *PH24_230 | 96.8 |
| PH24_275 | 92.5 |
| AR-AC | 95.1 |

Note: *PH24_230 hydrochar with the higher percent removal of lead outperforming AR-AC

| Temperature ($^0$C) | Time (Hours) | $k_2$ (g/mg/min) | Ho (mg/g/min) | $R^2$ | $q_e$ (calculated) (mg/g) |
|---|---|---|---|---|---|
| 180 | 6 | 0.122 | 2.119 | 0.99995 | 4.167 |
| | 12 | 1.002 | 21.787 | 0.99940 | 4,664 |
| | 24 | 4.025 | 82.645 | 0.99998 | 4,531 |
| 230 | 6 | 0.975 | 22.124 | 0.99994 | 4.764 |
| | 12 | 0.361 | 7.402 | 0.99969 | 4.527 |
| | 24 | 3.277 | 76.923 | 1 | 4.844 |
| 275 | 6 | 0.068 | 1.618 | 0.99383 | 4.880 |
| | 12 | 0.288 | 6.289 | 0.99945 | 4.675 |
| | 24 | 0.047 | 1.094 | 0.98427 | 4.849 |
| | AR-AC | 2.946 | 66.667 | 0.99998 | 4.757 |

**Table 6:** Pseudo-second order kinetic parameters for the adsorption of $Pb^2$

**Supplemental Data**

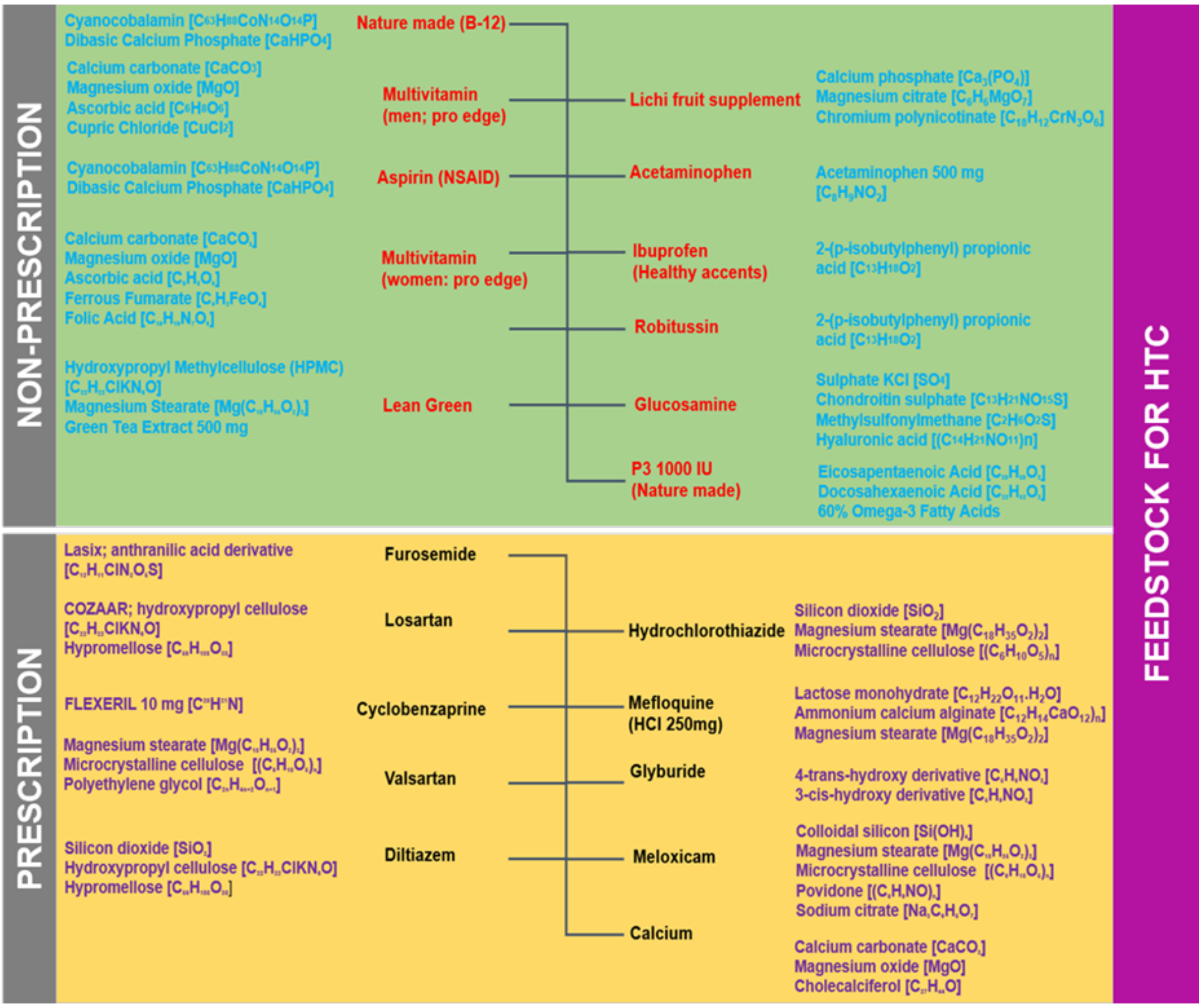

**Fig. S1.** The composition of waste pharmaceutical feedstock made up of both prescription and over the counter medicines.

**Table S1.** Types of pharmaceuticals, active ingredients, weight and weight percent

| Name | Active ingredients/Chemical Structure | Weight (g) | % weight | Prescri/non-prescription |
|---|---|---|---|---|
| Lichi fruit supplement | Calcium phosphate [$Ca_3(PO_4)$]<br>Magnesium citrate [$C_6H_6MgO_7$]<br>Chromium polynicotinate [$C_{18}H_{12}CrN_3O_6$] | 88.9 | 7.7 | Non-prescription |
| Hydrochlorothiazide | Silicon dioxide [$SiO_2$]<br>Magnesium stearate [$Mg(C_{18}H_{35}O_2)_2$]<br>Microcrystalline cellulose [$(C_6H_{10}O_5)_n$] | 5.7 | 0.49 | Prescription |
| Acetaminophen (Healthy ingredients) | Acetaminophen 500 mg [$C_8H_9NO_2$] | 56.1 | 4.86 | Non-prescription |
| Ibuprofen (Healthy accents) | 2-(p-isobutylphenyl) propionic acid [$C_{13}H_{18}O_2$] | 32.1 | 2.78 | Non-prescription |
| Robitussin | Dextromethorphan HBr [$C_{18}H_{25}NO$]<br>Guaifenesin [$C_{10}H_{14}O_4$] | 12.5 | 1.08 | Non-prescription |
| Nature made (B-12) | Cyanocobalamin [$C_{63}H_{88}CoN_{14}O_{14}P$]<br>Dibasic Calcium Phosphate [$CaHPO_4$] | 91.0 | 7.88 | Non-prescription |
| Glucosamine (Joint Therapy) | Sulphate KCl [$SO_4$]<br>Chondroitin sulphate [$C_{13}H_{21}NO_{15}S$]<br>Methylsulfonylmethane [$C_2H_6O_2S$]<br>Hyaluronic acid [$(C_{14}H_{21}NO_{11})_n$] | 26.3 | 2.28 | Non-prescription |
| Ibuprofen (assured) | 2-(p-isobutylphenyl) propionic acid [$C_{13}H_{18}O_2$] | 11.6 | 1.00 | Non-prescription |
| Mefloquine (HCl 250mg) | Lactose monohydrate [$C_{12}H_{22}O_{11}.H_2O$]<br>Ammonium calcium alginate [$C_{12}H_{14}CaO_{12})_n$] | 2.0 | 0.17 | Prescription |
| | Magnesium stearate [$Mg(C_{18}H_{35}O_2)_2$] | | | |
| P3 1000 IU<br>(Nature made) | Eicosapentaenoic Acid [$C_{20}H_{30}O_2$]<br>Docosahexaenoic Acid [$C_{22}H_{32}O_2$]<br>60% Omega-3 Fatty Acids | 29.2 | 2.53 | Non-<br>prescription |
| Mucinex | Dextromethorphan HBr [$C_{18}H_{25}NO$]<br>Guaifenesin [$C_{10}H_{14}O_4$] | 26.4 | 2.29 | Non-<br>prescription |
| Meloxicam | Colloidal silicon [$Si(OH)_4$]<br>Magnesium stearate [$Mg(C_{18}H_{35}O_2)_2$]<br>Microcrystalline cellulose [$(C_6H_{10}O_5)_n$]<br>Povidone [$(C_6H_9NO)_n$]<br>Sodium citrate [$Na_3C_6H_5O_7$] | 3.9 | 0.33 | Prescription |
| Acetaminophen<br>(equate) | Acetaminophen 500 mg [$C_8H_9NO_2$] | 59.1 | 5.12 | Non-<br>prescription |
| Glyburide | 4-trans-hydroxy derivative [$C_5H_9NO_3$]<br>3-cis-hydroxy derivative [$C_5H_9NO_3$] | 17.4 | 1.51 | Prescription |
| Furosemide | Lasix; anthranilic acid derivative [$C_{12}H_{11}ClN_2O_5S$] | 7.3 | 0.63 | Prescription |
| Losartan | COZAAR; hydroxypropyl cellulose [$C_{22}H_{22}ClKN_6O$]<br>Hypromellose [$C_{56}H_{108}O_{30}$] | 1.9 | 0.17 | Prescription |
| Cyclobenzaprine | FLEXERIL 10 mg [$C_{20}H_{21}N$] | 0.9 | 0.08 | Prescription |
| Valsartan | -Magnesium stearate [$Mg(C_{18}H_{35}O_2)_2$]<br>-Microcrystalline cellulose [$(C_6H_{10}O_5)_n$]<br>-Polyethylene glycol [$C_{2n}H_{4n+2}O_{n+1}$] | 18.3 | 1.59 | Prescription |
| Diltiazem | Silicon dioxide [$SiO_2$]<br>Hydroxypropyl cellulose [$C_{22}H_{22}ClKN_6O$] | 103.5 | 8.97 | Prescription |
| | Hypromellose [$C_{56}H_{108}O_{30}$] | | | |
| Lean Green | Hydroxypropyl Methylcellulose (HPMC) [$C_{22}H_{22}ClKN_6O$]<br>Magnesium Stearate [$Mg(C_{18}H_{35}O_2)_2$]<br>Green Tea Extract 500 mg. | 40.1 | 3.47 | Non-prescription |
| Multivitamin (men; pro edge) | Calcium carbonate [$CaCO_3$]<br>Magnesium oxide [$MgO$]<br>Ascorbic acid [$C_6H_8O_6$]<br>Cupric Chloride [$CuCl_2$] | 41.4 | 3.59 | Non-prescription |
| Aspirin (NSAID) | Aspirin (NSAID) 325 mg [$C_9H_8O_4$] | 51.5 | 4.46 | Non-prescription |
| Multivitamin (women: pro edge) | Calcium carbonate [$CaCO_3$]<br>Magnesium oxide [$MgO$]<br>Ascorbic acid [$C_6H_8O_6$]<br>Ferrous Fumarate [$C_4H_2FeO_4$]<br>Folic Acid [$C_{19}H_{19}N_7O_6$] | 81 | 7.02 | Non-prescription |
| Calcium | Calcium carbonate [$CaCO_3$]<br>Magnesium oxide [$MgO$]<br>Cholecalciferol [$C_{27}H_{44}O$] | 346.4 | 30.0 | Prescription |
| Total | | 1154.5 | 100 | |